\documentclass[preprint,superscriptaddress]{revtex4}
\usepackage[T1]{fontenc}
\usepackage[english]{babel}
\usepackage{array,multirow} 
\usepackage{amsfonts,amsmath,amssymb,bm}
\usepackage{graphicx}
\usepackage{color}
\usepackage{url,hyperref}
\hypersetup{colorlinks,allcolors=blue}
\usepackage{caption}
\usepackage{subcaption}
\newcommand{\vD}{v_{\rm D}}
\newcommand{\vSM}{v_{\rm SM}}
\newcommand{\vM}{v_{\rm M}}

\begin{document}
\title{Constraints on energy scales from dark matter decay in a gauged $B-L$ model}
\author{Guillermo Gambini}
\affiliation{Instituto de F\'isica Gleb Wataghin, UNICAMP, Campinas, SP, Brasil}
\affiliation{McGill University, Department of Physics, 3600 University Street, Montr\'eal, QC H3A 2T8, Canada}
\author{Pedro C. de Holanda}
\affiliation{Instituto de F\'isica Gleb Wataghin, UNICAMP, Campinas, SP, Brasil}
\author{Saulo Carneiro}
\affiliation{Instituto de F\'isica, Universidade Federal da Bahia, 40210-340, Salvador, BA, Brasil}

\begin{abstract}
Popular extensions of the standard model of particle physics feature new fields and symmetries which could, for example, dynamically generate neutrino masses from $B-L$ spontaneous symmetry breaking. If a new light scalar that decays into dark radiation appears in the spectrum of the theory, it could significantly modify the cosmological observables. In this case, cold dark matter could have a stable and a decaying component and limits on its decay rate $\Gamma_{\rm dcdm}$ can be used to put constraints on the new energy scales of a given model. We illustrate this idea using a gauged $B-L$ model where the dark radiation is in the form of light neutrinos.
\end{abstract}

\maketitle	
\thispagestyle{empty}

\section{Introduction\label{introduction}}
One of the driving forces of intensive research nowadays is the nature of the dark sector and its interactions with the content of the standard model of particle physics (SM). This sector is embedded in the cosmological standard model ($\Lambda$CDM) in the form of dark matter (DM) and dark energy. While the latter accounts for the accelerated expansion of the universe and its effects manifest in the form of a cosmological constant $\Lambda$, the former is most likely a new yet-unknown form of matter. The large amount of astronomical and cosmological observations are compatible with a cold relic whose self-interaction could solve, for example, small-scale problems \cite{Kaplinghat:2015aga,Tulin:2017ara,McDermott:2017vyk,Cline:2020gon}, while its interactions with SM particles could explain anomalies in terrestrial experiments \cite{Lindner:2020kko,Ge:2021cjz,Ge:2021snv,Cline:2019snp,Cline:2020mdt}, as well as discrepancies between low-redshift and cosmic microwave background (CMB) determinations of the Hubble parameter $H_0$ and the matter perturbation amplitude $\sigma_8$ \cite{Sam,Alcaniz:2019kah,Blinov:2020uvz,Abellan:2020pmw, Escudero:2021rfi}.

The resolution of these $H_0$ and $\sigma_8$ tensions \cite{Riess:2011yx,Heymans:2013fya,Planck:2015lwi,Riess:2016jrr} could also indicate the presence of a new neutrino species \cite{Carneiro:2018xwq}, primordial black holes \cite{Nesseris:2019fwr}, decaying dark matter \cite{Enqvist:2015ara} and more (see \cite{DiValentino:2021izs} for a review). In particular, a natural (decaying) DM candidate is the Majoron which  appears in models where the $B-L$ symmetry is broken by gravitational effects \cite{Akhmedov:1992et} or spontaneously broken by the vacuum expectation value (VEV) of a new complex singlet scalar \cite{Chikashige:1980ui}. Neutrino masses could be a consequence of this spontaneous symmetry breaking (SSB) at high scales if neutrinos are Majorana particles. Besides that, tiny Dirac neutrino masses can be achieved by coupling new right-handed neutral fermions and active neutrinos to new scalar doublets with VEVs $v_{\rm D} \ll v_{\rm SM} \approx$ 246 GeV \cite{Davidson:2009ha}. This dynamical generation of Dirac neutrino masses can be extended to Majorana neutrino masses by introducing new scalar singlets with larger VEVs $v_{\rm SM} \ll v_{\rm M}$ \cite{Montero:2011jk}. In this work, we put constraints on these Dirac and Majorana VEVs from the cosmic evolution and decay of a massive Majoron DM candidate that arises in gauged $B-L$ models and is able to constitute all or a fraction of the DM respecting limits from the cosmic microwave background (CMB) and the large-scale structure (LSS) of the universe.

In section 2 we briefly describe the gauged $B-L$ model we use to illustrate the constraints cosmology puts on these new scales $v_D$ and $v_M$. Constraints from late-time neutrino production and observational limits are presented in sections 3 and 4, respectively. We conclude in section 5.

\section{A $B$--$L$ model with exotic charges}
The model we use to study decaying DM is based on the gauge group $SU(3)_{\rm C} \otimes SU(2)_{\rm L} \otimes U(1)_{\rm Y'} \otimes U(1)_{\rm B-L}$, where C, L, Y$'$, and $B$--$L$ stand for color, left chirality, a new charge different from SM hypercharge, and baryon minus lepton number, respectively. Taking $B$--$L$ as a gauge symmetry, many models arise from the set of solutions to anomaly equations   \cite{Costa:2019zzy}. In Ref.\cite{montero2009gauging}, a model with three right-handed neutrinos with $B$--$L=5,-4,-4$ charges is proposed. Its representation content is that of the SM slightly extended by these new particles and six new scalars: two doublets $\Phi_{1,2}$ and four singlets $\phi_{1,2,3,X}$,
\begin{equation}
    SU(2)_{\rm L} \otimes U(1)_{\rm Y'} \otimes U(1)_{\rm B-L} \xrightarrow[]{\left< \phi \right>} SU(2)_{\rm L} \otimes U(1)_{\rm Y} \xrightarrow[]{\left< H,\Phi \right>} U_{\rm em}.
\end{equation}
All of these new scalars have unique $B$--$L$ quantum numbers \cite{sanchez2014complex}.

In this model, the most general renormalizable and gauge-invariant scalar potential reads
\begin{eqnarray}
V_{B-L} &=& -\mu^2_H H^{\dagger}H + \lambda_H |H^{\dagger}H |^2  \nonumber \\
&& - \mu_{11}^2 \Phi_1^{\dagger}\Phi_1 + \lambda_{11} |\Phi_1^{\dagger}\Phi_1|^2 - \mu_{22}^2 \Phi_2^{\dagger}\Phi_2 + \lambda_{22} |\Phi_2^{\dagger}\Phi_2|^2 -\mu_{s\alpha}^2 |\phi_{\alpha}|^2 +\lambda_{s\alpha} |\phi_{\alpha}^{*} \phi_{\alpha}|^2 \nonumber \\
&& + \lambda_{12} |\Phi_1|^2 |\Phi_2|^2 + \Lambda_{H\gamma} |H|^2 |\Phi_\gamma|^2 + \Lambda_{Hs\alpha} |H|^2 |\phi_\alpha|^2 + \Lambda^{'}_{\gamma \alpha} |\Phi_\gamma|^2 |\phi_\alpha|^2 + \Delta_{\alpha \beta} (\phi_\alpha^{*}\phi_\alpha)(\phi_\beta^{*}\phi_\beta) \nonumber \\
&& + \lambda_{12}^{'} (\Phi_1^{\dagger} \Phi_2) (\Phi_2^{\dagger} \Phi_1)  + \Lambda_{H\gamma}^{'} (H^{\dagger} \Phi_\gamma) (\Phi_\gamma^{\dagger} H) \nonumber \\
&& + [ \beta_{123} \phi_1 \phi_2 (\phi_3^{*})^2 + \Phi^{\dagger}_1 \Phi_2 (\beta_{13}\phi_1 \phi_3^{*} + \beta_{23}\phi_2^{*}\phi_3) - i\kappa_{H1X} \Phi_1^T \tau_2 H \phi_X \nonumber \\
&& - i\kappa_{H2 X} (\Phi_2^T \tau_2 H ) (\phi_X^{*})^2 + \beta_X (\phi_X^{*}\phi_1)(\phi_2\phi_3) + \beta_{3X}(\phi_X^{*}\phi_3^3) + H.c. ],
\nonumber
\label{VBL}
\end{eqnarray}
where $\gamma=1,2$, $\alpha,\beta=1,2,3,X$, and $\alpha>\beta$ in the $\Delta_{\alpha \beta}$ terms.
The quadratic potential reads $V_2=\frac{1}{2} \bm{\varphi}^\tau M \bm{\varphi}$, where the scalar mass matrix can be written as $M=\partial^2 V_{B-L}/\partial \bm{\varphi}^2$. Being particularly interested in the CP-odd sector, we take $\bm{\varphi}=\{\text{Im}_\varphi\}$ with $\varphi^0=\frac{1}{\sqrt{2}}(v_\varphi + \text{Re}_\varphi + \text{Im}_\varphi)$ and $\varphi = H,\Phi_1,\Phi_2,\phi_1,\phi_2,\phi_3,\phi_X$. In order to find the squared masses, we proceed to find the roots of the eigenvalue equation $\det(\text{M}-\lambda\, \text{I}_{7\times7}) = 0$. This is a hard task since the large number of free parameters forbids useful analytical expressions. For this reason, we depart from the `general model' approach (where all couplings are independent) and proceed to simplify the model assuming similar terms in $V_{B-L}$ may have similar couplings. Namely, 
\begin{eqnarray}
&&\lambda_{11}=\lambda_{12}=\lambda_{s1}=\lambda_{s3}=\lambda_{sX} \nonumber, \\
&&\Lambda_{H1}=\Lambda_{H1}=\Lambda_{H2}=\Lambda_{Hs1}=\Lambda_{Hs3}=\Lambda_{HsX}=\Lambda_{H1}'=\Lambda_{H2}'\nonumber, \\
&&\Lambda_{11}'=\Lambda_{13}'=\Lambda_{1X}'=\Lambda_{21}'=\Lambda_{23}'=\Lambda_{2X}'=\lambda_{12}=\lambda_{12}'=\Delta_{13}=\Delta_{1X}=\Delta_{3X}\nonumber, \\
&&\Lambda_{12}'=\Lambda_{22}'=\Delta_{12}=\Delta_{23}=\Delta_{2X}.
\end{eqnarray}
We leave all other parameters free except for the VEVs of the new scalars which we make $v_{\rm D}$ for $\left<\Phi_i\right>$ and $v_{\rm M}$ for $\left<\phi_j\right>$, where $i=1,2$, $j=1,2,3,X$, and $v_{\rm D} \ll v_{\rm SM} \approx 246 \text{ GeV} \ll v_{\rm M}$. This choice was made in Ref.\cite{sanchez2014complex} in order to obtain significant phenomenological results and here we follow their work closely. In that paper, the masses in the CP-odd scalar sector were found in the limit where $\vD /\vM \rightarrow 0$,
\begin{equation}
    m_{I_1}^2 = 0, \hspace{0.1in} m_{I_2}^2 = 0, \hspace{0.1in} m_{I_3}^2 = \mathcal{O}(\vD/\vM),
    \label{mI123}
\end{equation}
\begin{equation}
    m_{I_4}^2 = \frac{1}{2} \left[ \Lambda_{Hs2} \vSM^2 + (\Lambda_{12}'+\lambda_{22}' - 2\Lambda_{Hs2} ) \vD^2 + (\Delta_{12}+\Delta_{23}+\Delta_{2X}) \vM^2 - 2\mu_{s2}^2   \right],
\end{equation}
\begin{equation}
    m_{I_{5,6}}^2 = \frac{1}{4} \vM \left[ (1+\sqrt{2})\vSM -2\beta_{13} \vM \mp \sqrt{4 \beta_{13}^2 \vM^2 +(3-2\sqrt{2}) \vSM^2 } \right],
\end{equation}
\begin{equation}
    m_{I_7}^2 = -5\beta_{3X} \vM^2, 
\end{equation}
where $I_i$ ($i=1...7$) are linearly-independent eigenvectors of M. Note that even if an analytical expression for $m_{I_3}^2$ is not shown in Eq.(\ref{mI123}), apart from the two Goldstone bosons, the remaining particles are much heavier than $I_3$ unless their masses are fine-tuned not to be so. 

A method to obtain approximate expressions for eigenvalues and eigenvectors is Rayleigh-Schr\"{o}dinger perturbation theory \cite{lancaster1985theory,mccartin2009rayleigh} where the mass matrix is expanded in terms of a small parameter \cite{Alvarez-Salazar:2019cxw}. In our case, $\zeta \equiv v_{\rm D}/v_{\rm M} \ll $1 is the small parameter that allows us to obtain the zeroth-order eigenvector (see Appendix \ref{appendix:Rayleigh}), 
\begin{equation}
    I_3^{(0)} = \frac{1}{\sqrt{10}} \text{Im}_{\phi_3} + \frac{3}{\sqrt{10}} \text{Im}_{\phi_X}.
\end{equation}
For the mass of $I_3$, we have found a better approximation using a different method (see Appendix \ref{appendix:mI3}). We defined three small parameters and found the lightest root of the characteristic polynomial of the scalar mass matrix. The mass of the lightest CP-odd scalar $I_3$ has been found to be a function of the two SSB scales and two dimensionless parameters,  $m^2_{I_3}=m^2_{I_3}(\vD, \vM, \beta_{13}, \beta_{3X})$. Namely, 
\begin{equation}
m_{I_3}^2 \simeq \frac{37\, \vSM \, \vM^2 \vD^2 \, \beta_{13}\, \beta_{3X}}{5 \sqrt{2}(1+\sqrt{2}) \vM^3 \beta_{13} \beta_{3X} - \vSM \left( 11 \vD^2 \beta_{13} + 10 \vM^2 \beta_{3X} \right)  },
\label{mI3new}
\end{equation}
where $\beta_{13}<0$ and $\beta_{3X}<0$ because of the positivity of other two masses. Numerically, as $v_{\rm SM} \approx 246$ GeV and $v_{\rm SM} \ll \vM$, for $\beta_{13}\sim \mathcal{O}$(1), we have
\begin{equation}
    m_{I_3} \approx 0.73 \sqrt{\frac{\vD}{1\text{ MeV}}}\sqrt{\frac{\vD}{\vM} }  \text{ GeV}.
    \label{eq:mI3}
\end{equation}
The unstable $I_3$ pseudoscalar decays mainly into active neutrinos, with decaying rate \cite{schechter1982neutrino}
\begin{equation}
\Gamma_{I_3 \rightarrow \nu \nu}\approx \frac{m_{I_3}}{16\pi} \frac{\sum_i m^2_{\nu_i}}{\vM^2},
\label{eq:decayrate}
\end{equation}
where $\vM$ is the SSB scale of the $B-L$ symmetry. Since neutrino masses are unknown, we take the sum of these squared masses as a free parameter (see Appendix \ref{appendix:sum2}). In this way, it is convenient to rewrite Eq.(\ref{eq:mI3}) and Eq.(\ref{eq:decayrate}) like
\begin{equation}
\frac{m_{I_3}}{\text{eV}} \approx 23.1 \left(\frac{\vD}{\text{KeV}} \right) \left( \frac{\text{1000 TeV}}{\vM} \right)^{1/2} ,
\label{eqmI3-b}
\end{equation}

\begin{equation}
\frac{\text{Gyr}}{\Gamma_{I_3\to\nu\nu}}=\frac{\tau_{I_3}}{\text{Gyr}} \approx 9.1 \left( \frac{5\times 10^{-3}\text{ eV}^2}{\sum m^2_\nu} \right) \left(\frac{\text{KeV}}{\vD} \right) \left( \frac{\vM}{\text{1000 TeV}} \right)^{5/2}.
\label{eqtau-b}
\end{equation}

\section{Late-time neutrino production}
Including the $I_3$ particle into the dynamics of the different components in the Universe, we can write the following equations:
\begin{eqnarray}
\dot{\rho}_{I_3} + 3 H \rho_{I_3} &=& -\Gamma \rho_{I_3},  \label{eq:rhoevol1}
\\
\dot{\rho}_{\nu}+4H\rho_{\nu}&=&+\Gamma\rho_{I_3}, \label{eq:rhoevol2}
\\
\dot{\rho}_{m} + 3 H \rho_{m} &=& 0,  
\label{eq:rhoevol3}
\end{eqnarray}
where $\rho_{I3}$ is the density of the unstable particle $I_3$, which produces an increase on the neutrino density $\rho_\nu$, and $\Gamma = \Gamma (I_3 \rightarrow \nu \nu)$. The stable cold dark matter is denoted by $\rho_m$.

Considering the I$_3$ freeze-out temperature T$_D$ $\gtrsim m_t \approx$ 173 GeV, the $I_3$ abundance can be written as \cite{Kolb:1990vq,lattanzi2007decaying}
\begin{equation}
\Omega_{I_3,0} \,h^2 = \frac{m_{I_3}}{1.25 \text{ KeV}} \exp \left( -t_0 / \tau \right),
\label{eq1:omegai3}
\end{equation}
where $\tau_{I_3}=1/\Gamma_{I_3 \rightarrow \nu \nu}$ is the $I_3$ lifetime and $t_0$ is the age of the Universe.
\begin{figure}
  \centering
  \includegraphics[width=.7\linewidth]{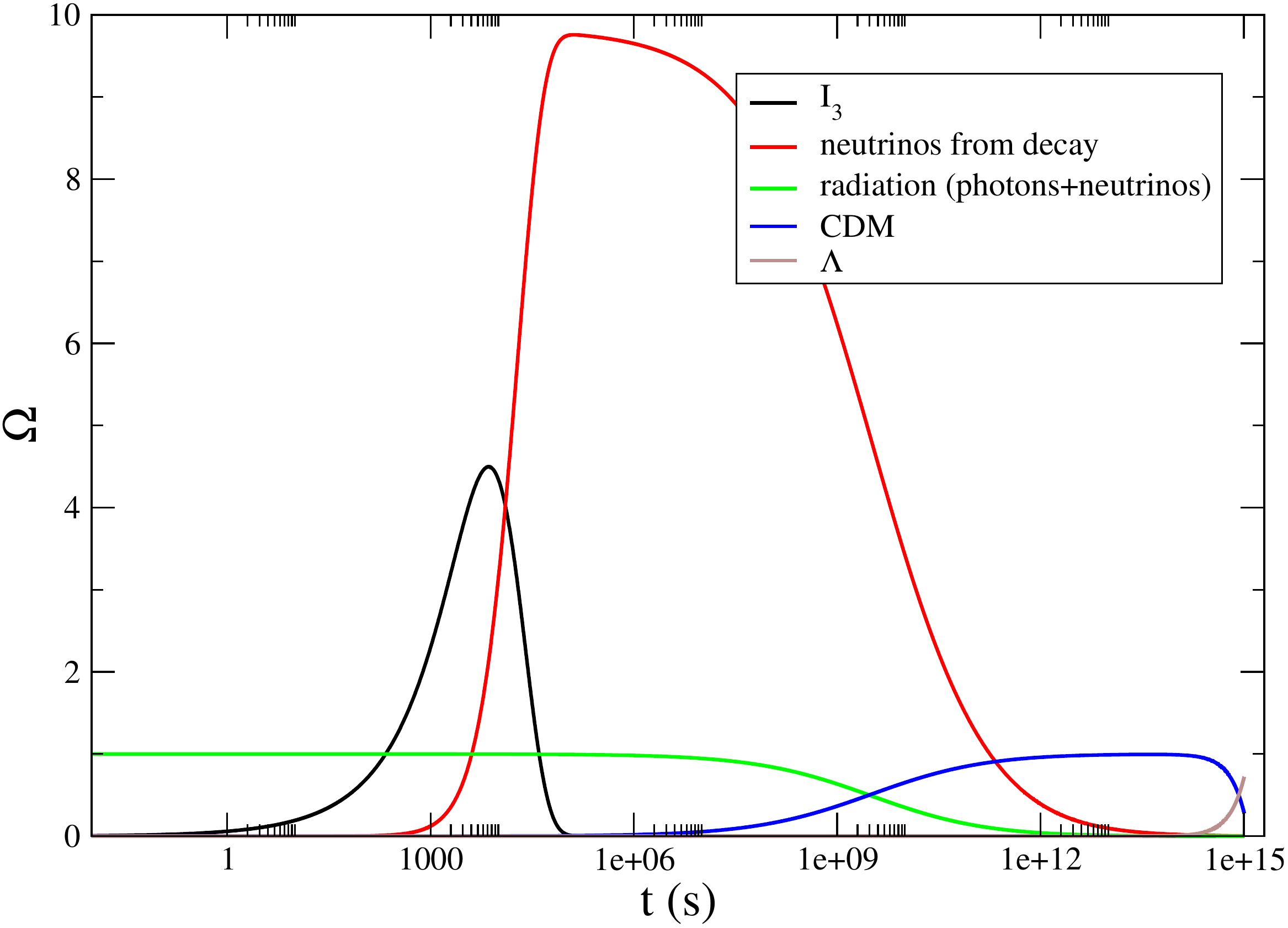}
  \caption{Relative densities for the case $\vD = 1$ MeV and $\vM = 1$ TeV.}
  \label{fig:densities}
\end{figure}
As we can see, both the lifetime and abundance of $I_3$ depend strongly on $\vD$ and $\vM$, so we will use these observables to put limits on these VEVs. As a first step, let's analyze the example $\vD = 1$ MeV and $\vM = 1$ TeV from \cite{sanchez2014complex}. These values can give valid neutrino masses by means of the type-I seesaw mechanism and also the correct relic abundance for a stable DM candidate when a $\mathbb{Z}_2$ symmetry is imposed on the $\phi_2$ singlet scalar. Using these values, we get
\begin{equation}
\tau_{I_3} \approx 0.1 \text{ year}~~~\text{and}~~~\Omega_{I_3,0} \, h^2 \approx 10^3\exp \left( -t_0 /(0.1 \text{ yr}) \right) .
\end{equation}
So, in this case $I_3$ represents a negligible fraction of the total dark matter in the universe today. This is fine if there is another DM candidate that can account for the cold dark matter (CDM) relic density $\Omega_{cdm,0}$. However, we can calculate the amount of produced neutrinos from the decays of $I_3$ and check whether this is in conflict with observations. This calculation is done numerically by integrating equations~(\ref{eq:rhoevol1}) to~(\ref{eq:rhoevol3}) with initial conditions for $\rho_{I_3}$ derived from eq.~(\ref{eq1:omegai3})  and $\rho_{\nu,i}=0$ for the neutrinos produced from $I_3$ decays.  

In Fig.~\ref{fig:densities} the fractional abundances are shown in comparison with the standard cosmological scenario. Due to the large abundance of $I_3$ at early times, there is a period of CDM dominance driven by $I_3$ around $t\sim 100$ s. When the decay starts to take place, the energetic content is populated by the relativistic neutrinos, which would be the main ingredient in the Universe until the CDM starts to preponderate again. Such scenario would affect dramatically the cosmological observables, and would hardly be acceptable by the latest perturbation data. Therefore, since some choices for the model parameters can affect the cosmology, we present in the following some weak limits on such parameters.

\section{Observational limits}
If $I_3$ is a long-lived particle ($\tau_{I_3} \gg t_0$) we have, from Eqs. (\ref{eqmI3-b}) and (\ref{eqtau-b}),
\begin{eqnarray}
  \Omega_{I_3} &\approx & \frac{m_{I_3}}{1.25 \text{ KeV}} h^{-2}, \nonumber \\ 
  &\approx & \frac{9\times 10^{-3}}{h^2} \left(\frac{\vD}{\text{KeV}} \right) \left( \frac{\text{1000 TeV}}{\vM} \right)^{1/2}, \nonumber \\
  &\approx & \frac{9\times 10^{-3}}{h^2} \left( \frac{\text{Gyr}}{\tau_{I_3}} \right)  \left( \frac{\vM}{\text{1000 TeV}} \right)^2,
\label{eq1:omegai3'}
\end{eqnarray}
assuming $\sum m_\nu^2=5\times 10^{-3} \text{ eV}^2$. This gives us an upper bound for the energy scale ($\vM$) of the $B-L$ spontaneous symmetry breaking, since the relative matter density is below 1,
\begin{equation}
    \vM \lesssim 6.9\, h \left( \frac{\tau_{I_3}}{\rm Gyr} \right)^{1/2} \times 10^3\ {\rm TeV}.
\end{equation}
For instance, if $\tau_{I_3}=100$ Gyr, using $h=0.674$ we get
\begin{equation}
    \vM \lesssim 46.6 \times 10^3\ {\rm TeV}.
\end{equation}
For lower values of $\tau_{I_3}$, we can make the following estimation. If at $t=\tau_{I_3}$ all $I_3$ is converted immediately to neutrinos, then the energy density would decrease with an extra scaling factor $a$ in comparison with the cold dark matter. Assuming a matter dominated universe, such event happens at
\[
a\approx \left(\frac {3H_0\tau_{I_3}}{2}\right)^{2/3}.
\]
Then,
\[
\Omega_{\nu} h^2 \approx (H_0\tau_{I_3})^{2/3}\frac{m_{I_3}}{1.25 \text{ KeV}} 
\approx (H_0\tau_{I_3})^{2/3}\,10^{-7}\left(\frac{1\text{ Gyr}}{\tau_{I_3}}\right)\left(\frac{\vM}{1 \text{ TeV}}\right)^2.
\]
Replacing $H_0\approx 0.1\,h$ Gyr$^{-1}$:
\[
\Omega_{\nu} h^2 \approx (0.1\,h)^{2/3}10^{-7}\left(\frac{1\text{ Gyr}}{\tau_{I_3}}\right)^{1/3}\left(\frac{\vM}{1 \text{ TeV}}\right)^2.
\]
For $\tau_{I_3}\approx 0.1$ yr, we would have $\vM< 76$ TeV, since $\Omega_{\nu} \lesssim 1$. These numbers help to restrict the space of parameters, however we would expect a much stronger restriction when analysing other cosmological observables. 
\begin{figure}[h]
  \centering
  \includegraphics[width=0.7\linewidth]{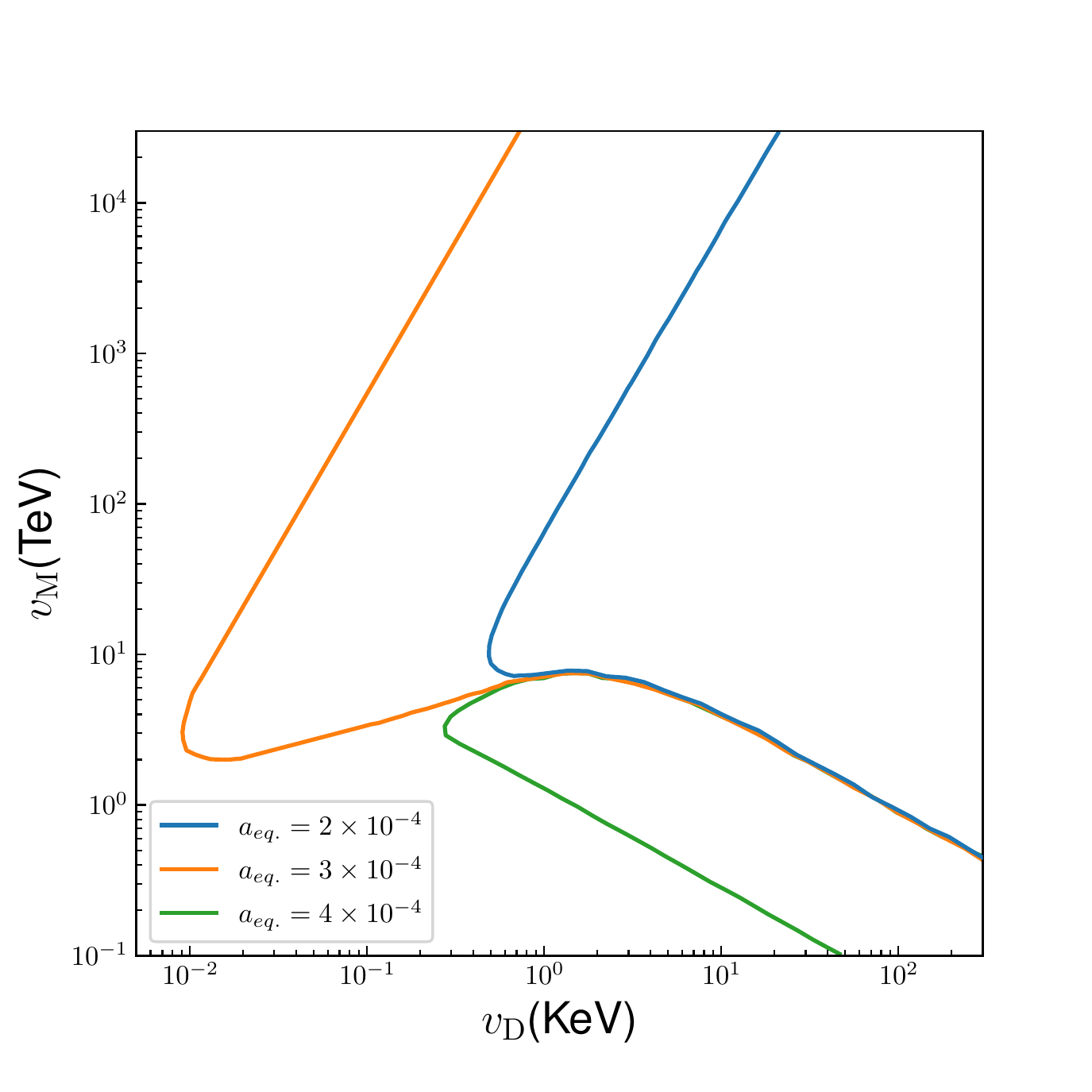} 
  \caption{Value of scale factor at matter-radiation equality. For reference, $a_{eq}\sim 3\times 10^{-4}$ is the $\Lambda$CDM prediction.}
  \label{fig:limits}
\end{figure}
For instance, we can calculate the scale factor when the radiation and matter densities are equal. A strong production of radiation (matter) would increase (decrease) the value of the scale factor, affecting the growth of structures or the CMB anisotropies. One way to estimate the limit on the  amount of radiation produced by the decay of $I_3$ would be to translate the limits on the number of neutrino families given by~\cite{Planck:2018vyg}, {\it i.e.} $N_{eff}=2.99\pm 0.17$, to a $3\sigma$ interval for the time of equality between radiation and non-relativistic species,
\begin{equation}
a_{eq.}= (3.0\pm 0.2)\times 10^{-4}.
\end{equation}
In Fig.~\ref{fig:limits} we show the scale factor at the equality. The safe choice would be to stay on the left of both blue and green lines. There is a fine-tuned region where the blue and green lines are almost indistinguishable, and it would produce the `correct' scale factor. 

\begin{figure}
        \centering
        \begin{subfigure}[b]{0.45\textwidth}
            \centering
            \includegraphics[width=\textwidth]{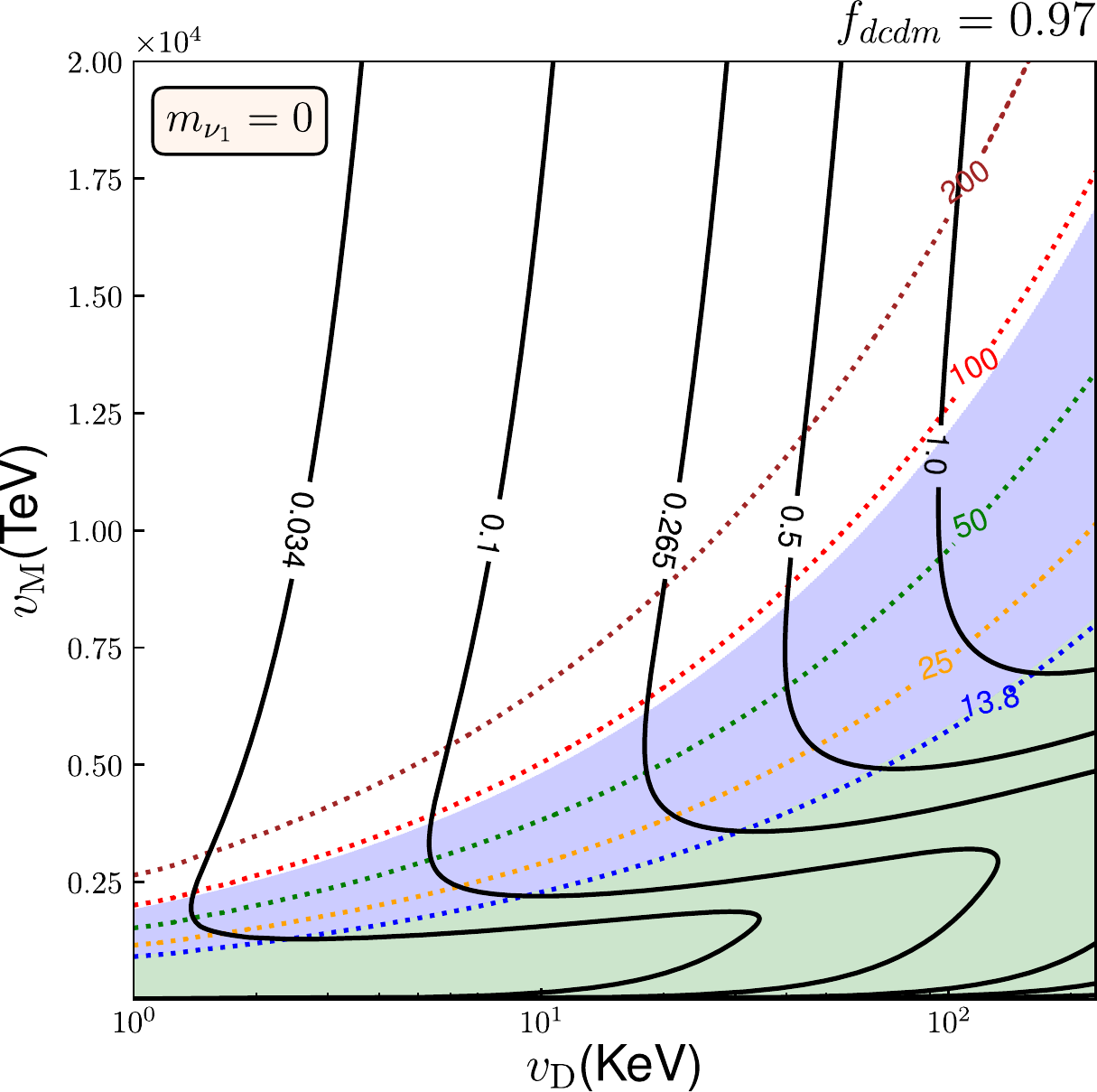}
        \end{subfigure}
        \hfill
        \begin{subfigure}[b]{0.45\textwidth}  
            \centering 
            \includegraphics[width=\textwidth]{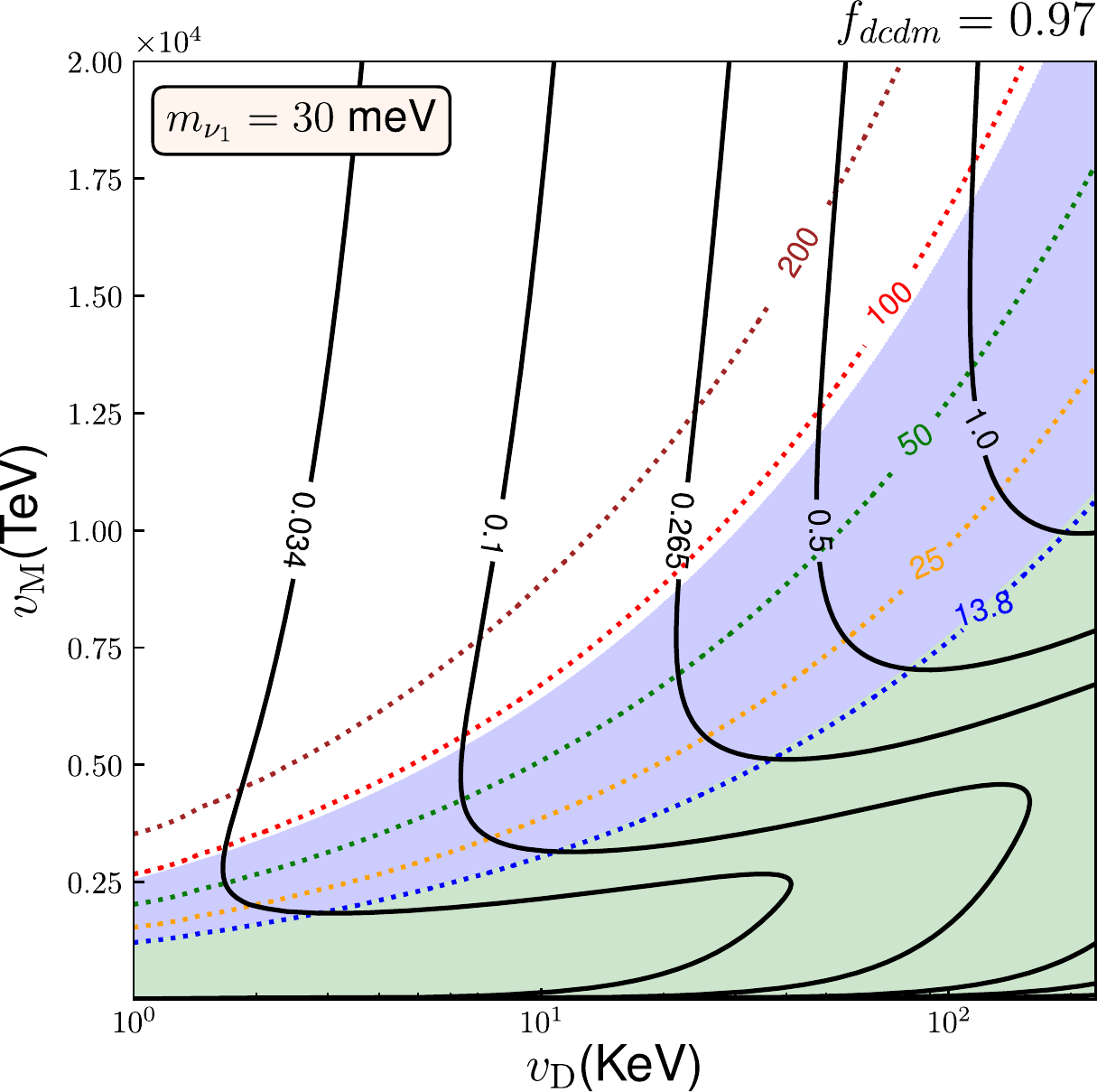}
        \end{subfigure}
        \begin{subfigure}[b]{0.45\textwidth}   
            \centering 
            \includegraphics[width=\textwidth]{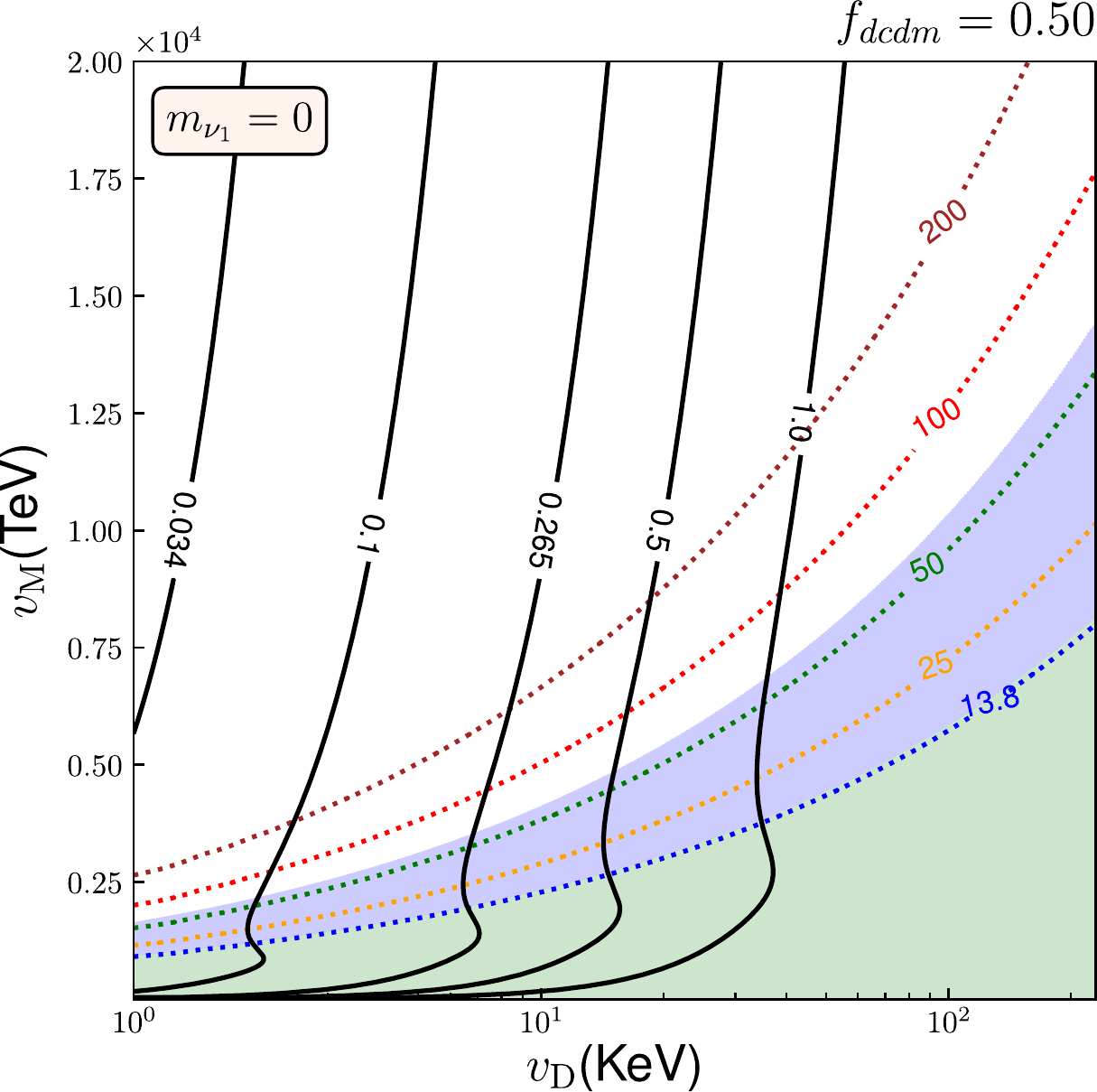}
        \end{subfigure}
        \hfill
        \begin{subfigure}[b]{0.45\textwidth}   
            \centering 
            \includegraphics[width=\textwidth]{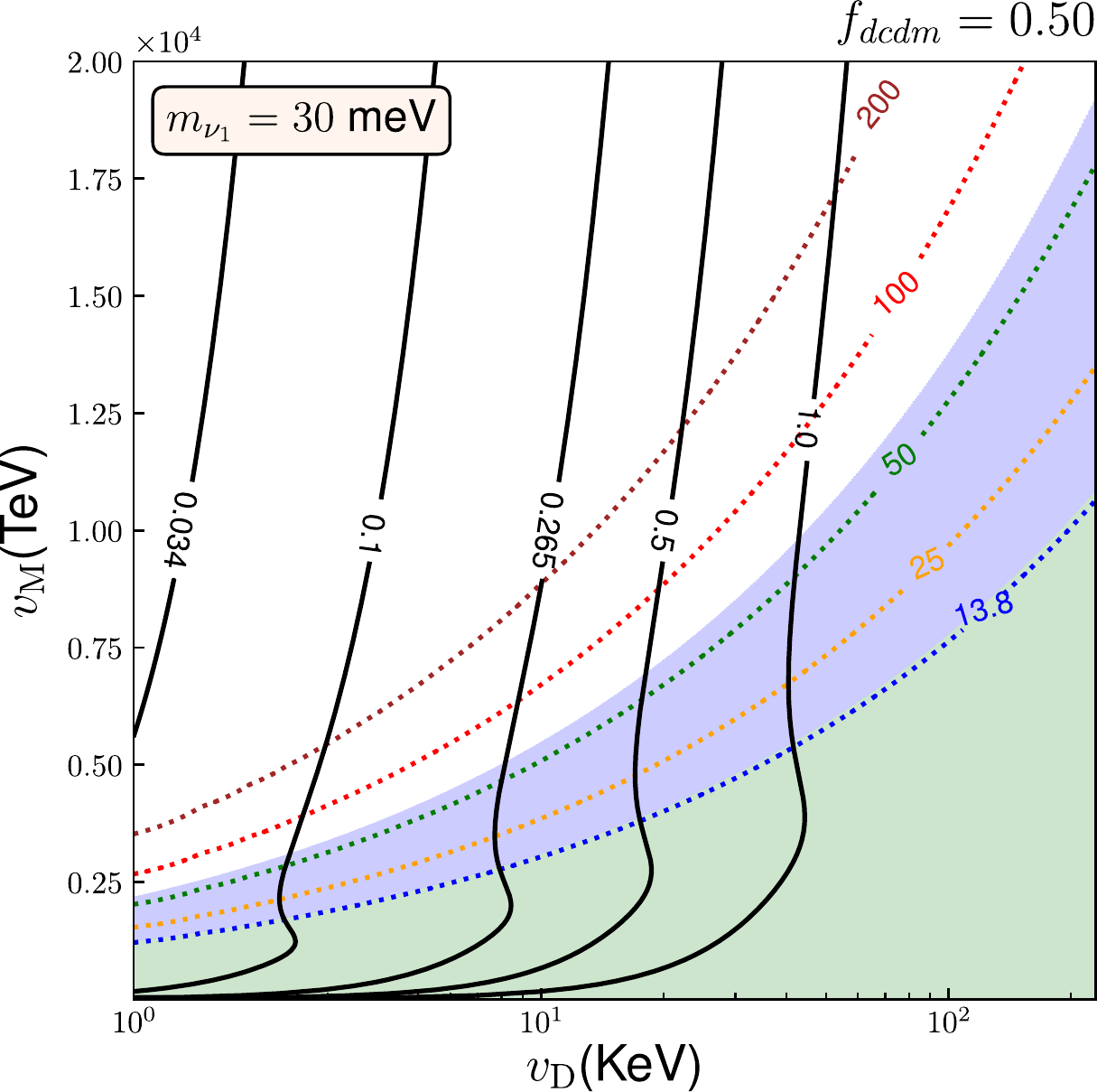}
        \end{subfigure}
        \begin{subfigure}[b]{0.45\textwidth}   
            \centering 
            \includegraphics[width=\textwidth]{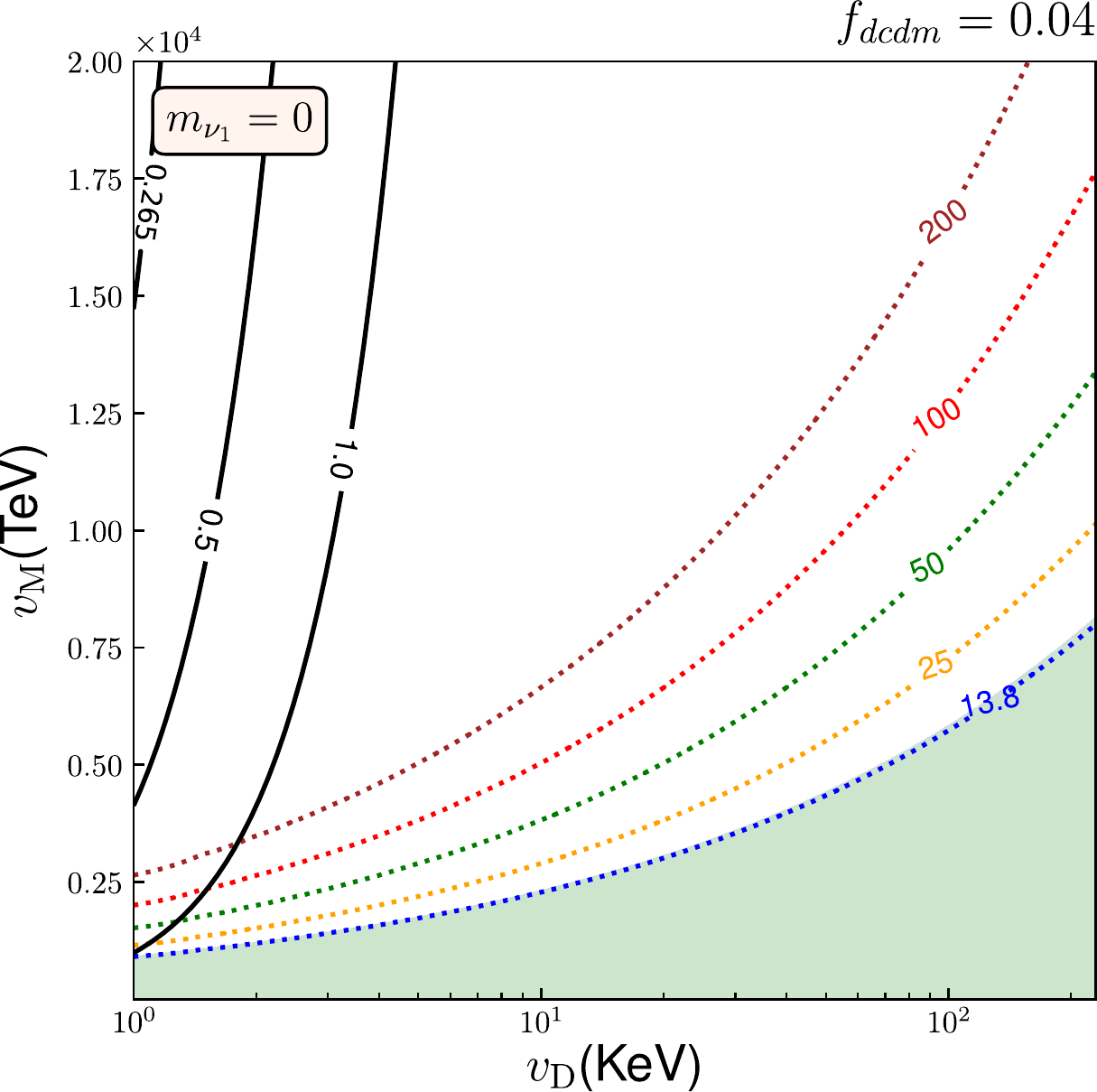}
        \end{subfigure}
        \hfill
        \begin{subfigure}[b]{0.45\textwidth}   
            \centering 
            \includegraphics[width=\textwidth]{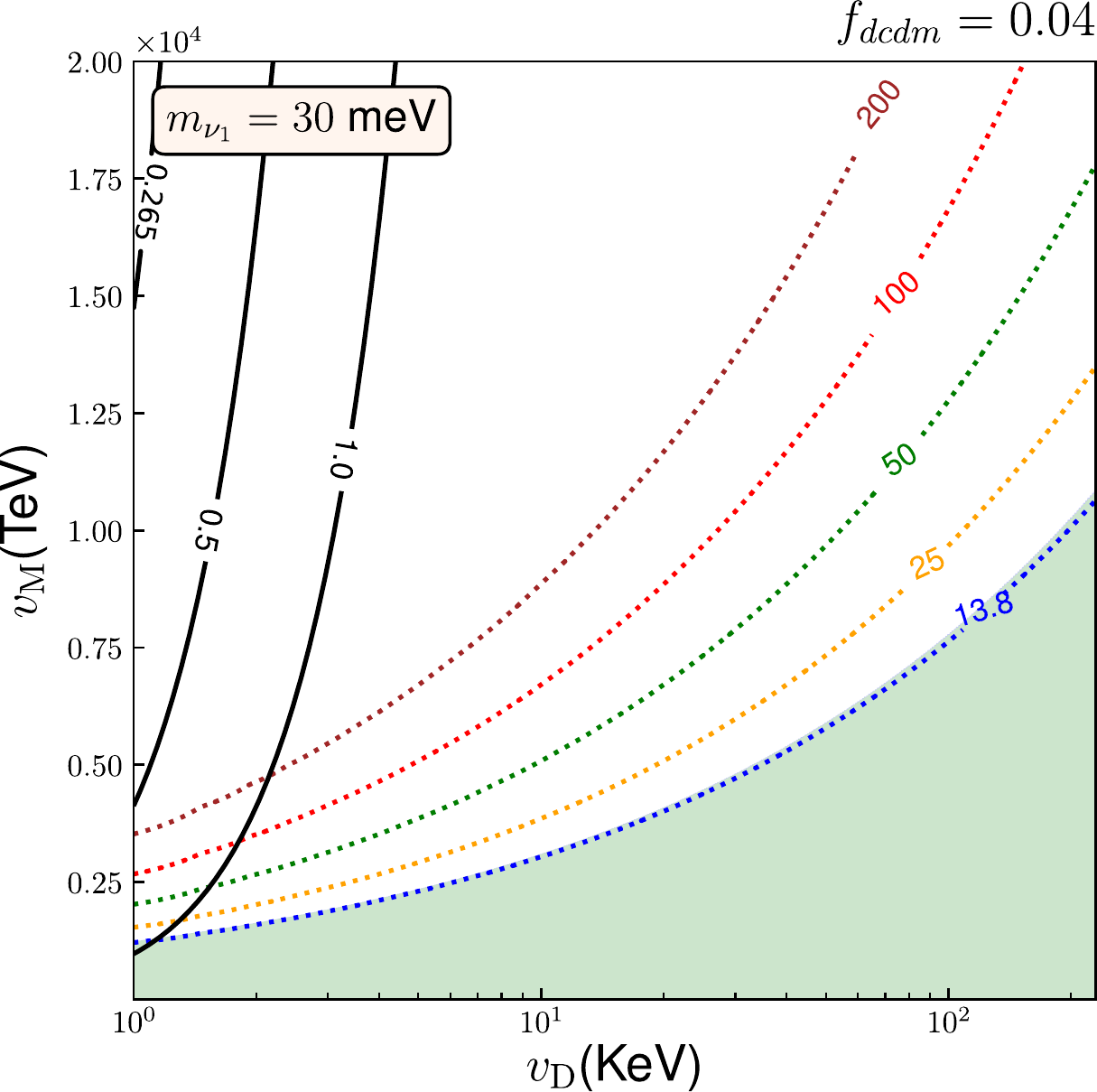}
        \end{subfigure}
        \caption[]{\footnotesize Normal hierarchy: DM relative density (solid black) and $I_3$ lifetime in Gyr (dotted) curves. Blue regions are ruled-out because $I_3$ represents too-large of a fraction of the cold dark matter and its mean lifetime is too short, even if it is larger than the present age of the universe. Green regions are excluded because $\Gamma(I_3 \to \nu\nu)>H_0$ and $f_{\rm dcdm}$ used in these plots are larger than 0.04.}
        \label{OmegalifetimeI3NH}
    \end{figure}

\begin{figure}
        \centering
        \begin{subfigure}[b]{0.45\textwidth}
            \centering
            \includegraphics[width=\textwidth]{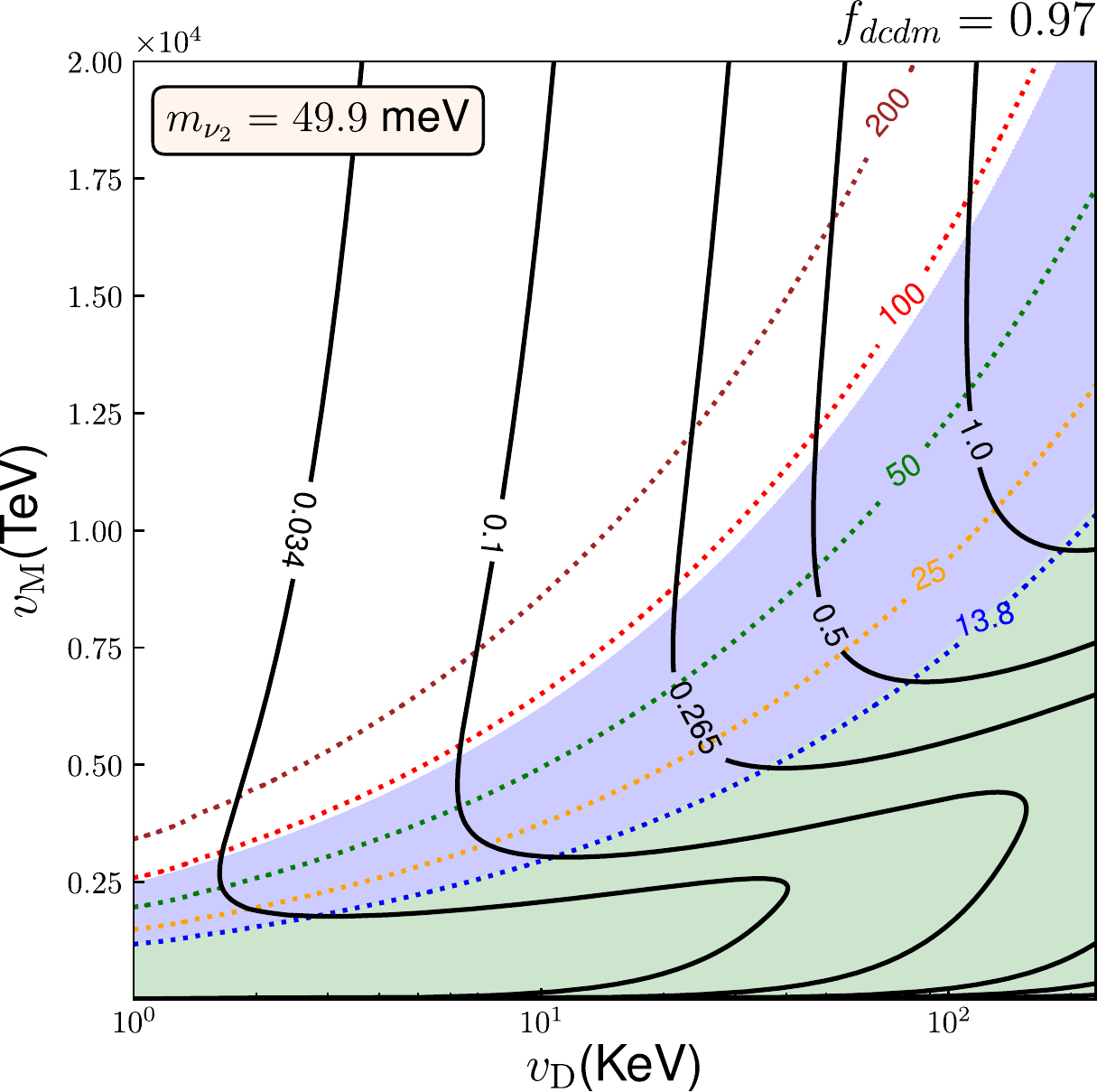}
        \end{subfigure}
        \hfill
        \begin{subfigure}[b]{0.45\textwidth}  
            \centering 
            \includegraphics[width=\textwidth]{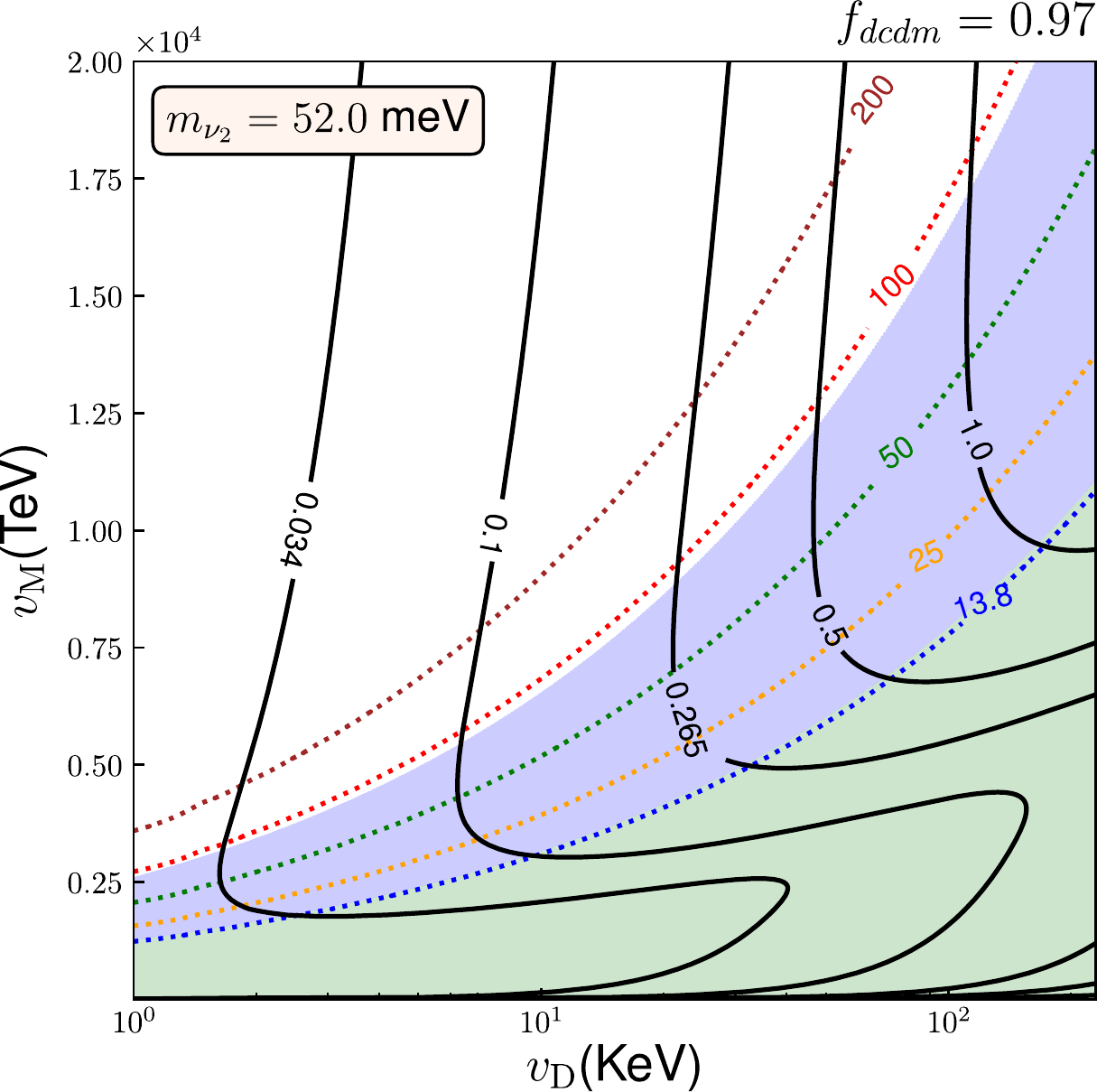}
        \end{subfigure}
        \begin{subfigure}[b]{0.45\textwidth}   
            \centering 
            \includegraphics[width=\textwidth]{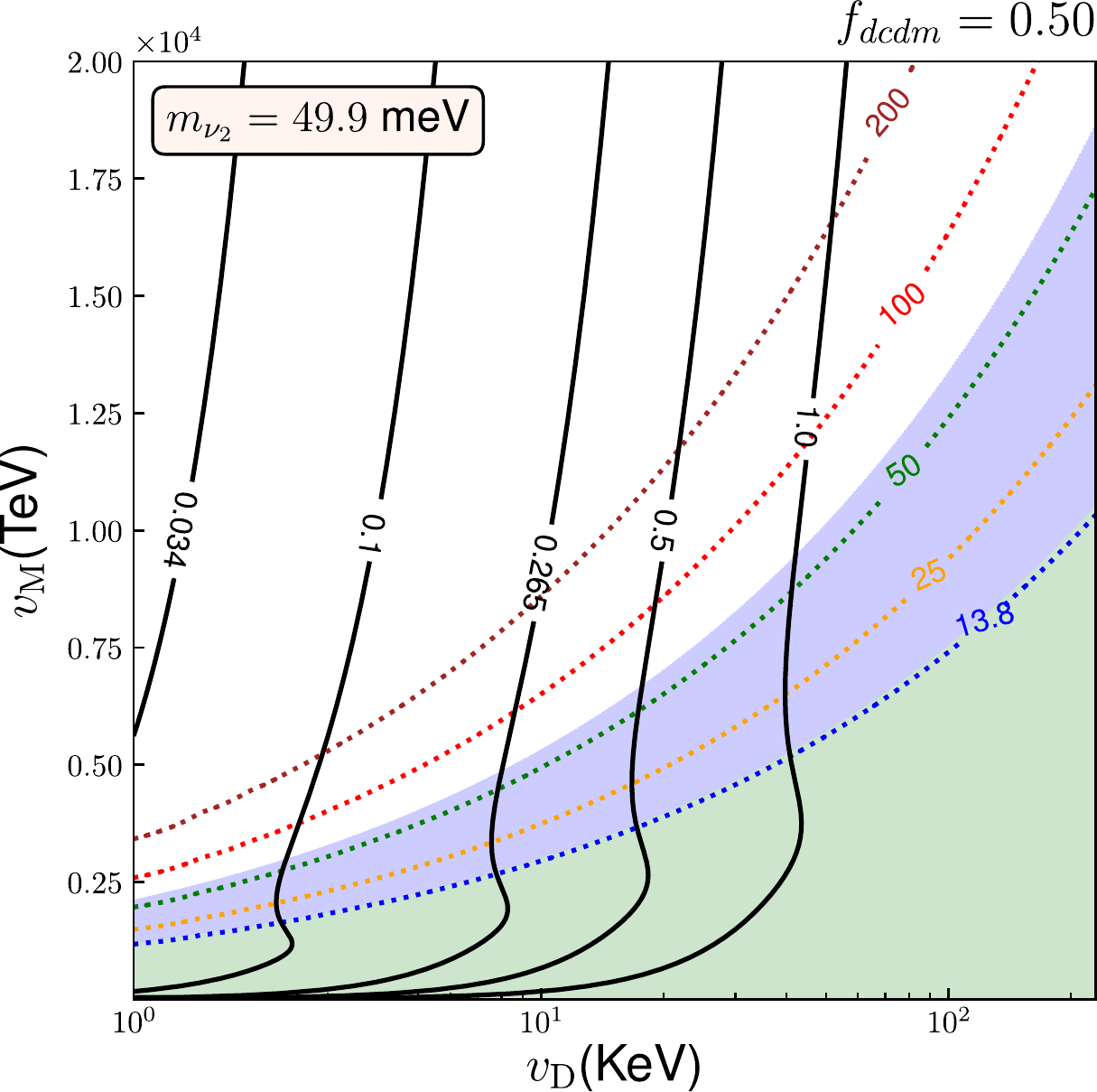}
        \end{subfigure}
        \hfill
        \begin{subfigure}[b]{0.45\textwidth}   
            \centering 
            \includegraphics[width=\textwidth]{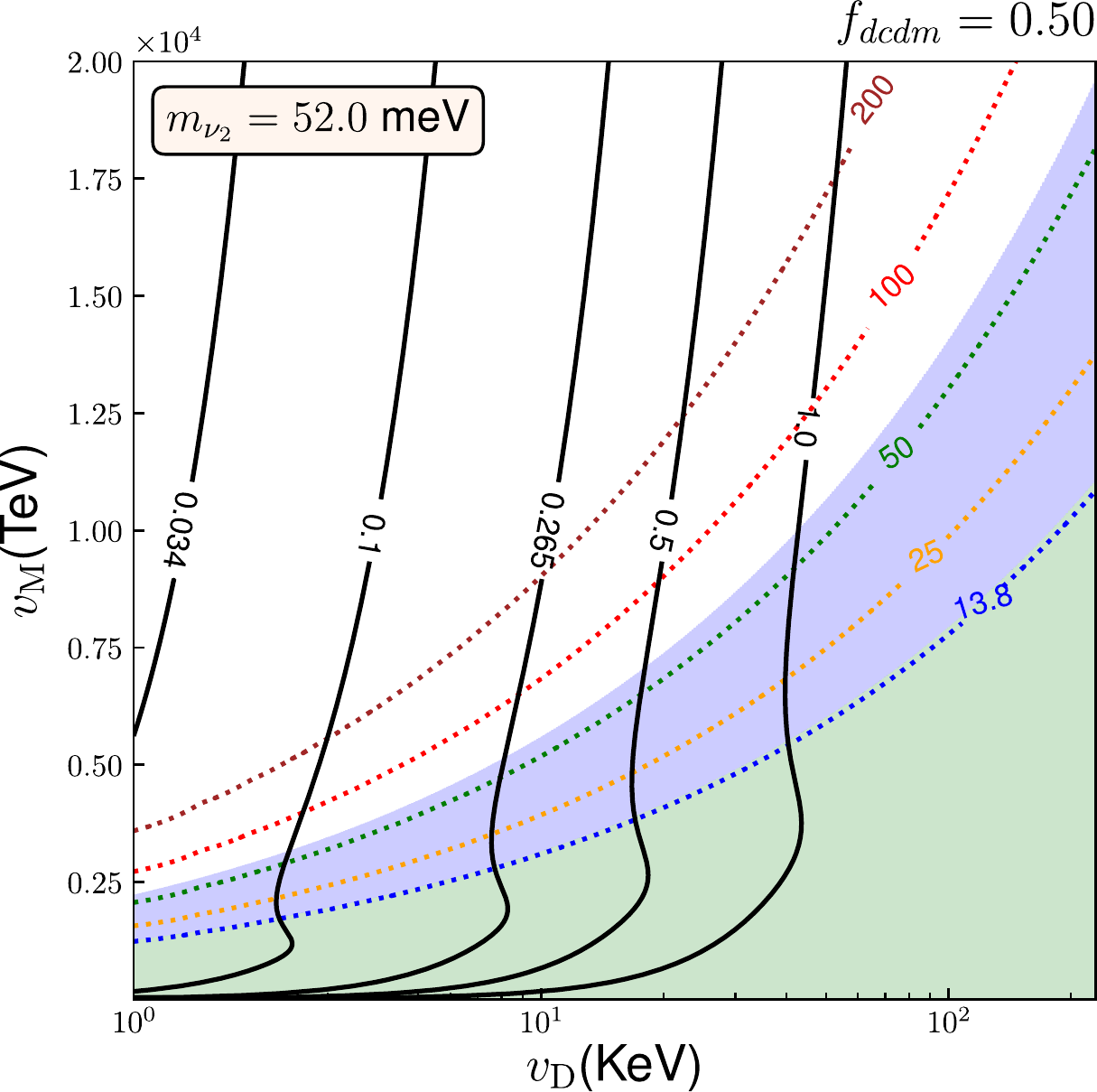}
        \end{subfigure}
        \begin{subfigure}[b]{0.45\textwidth}   
            \centering 
            \includegraphics[width=\textwidth]{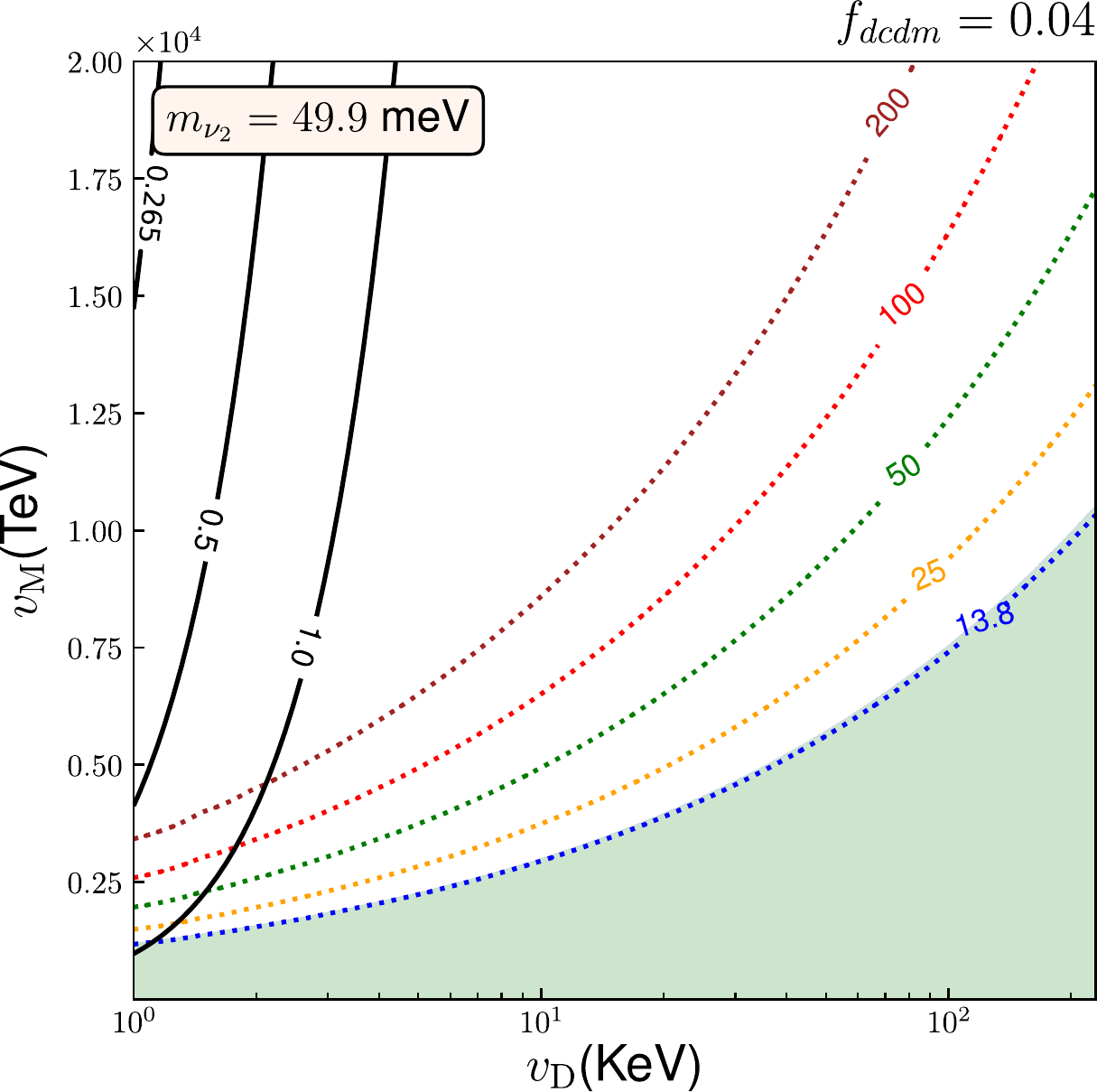}
        \end{subfigure}
        \hfill
        \begin{subfigure}[b]{0.45\textwidth}   
            \centering 
            \includegraphics[width=\textwidth]{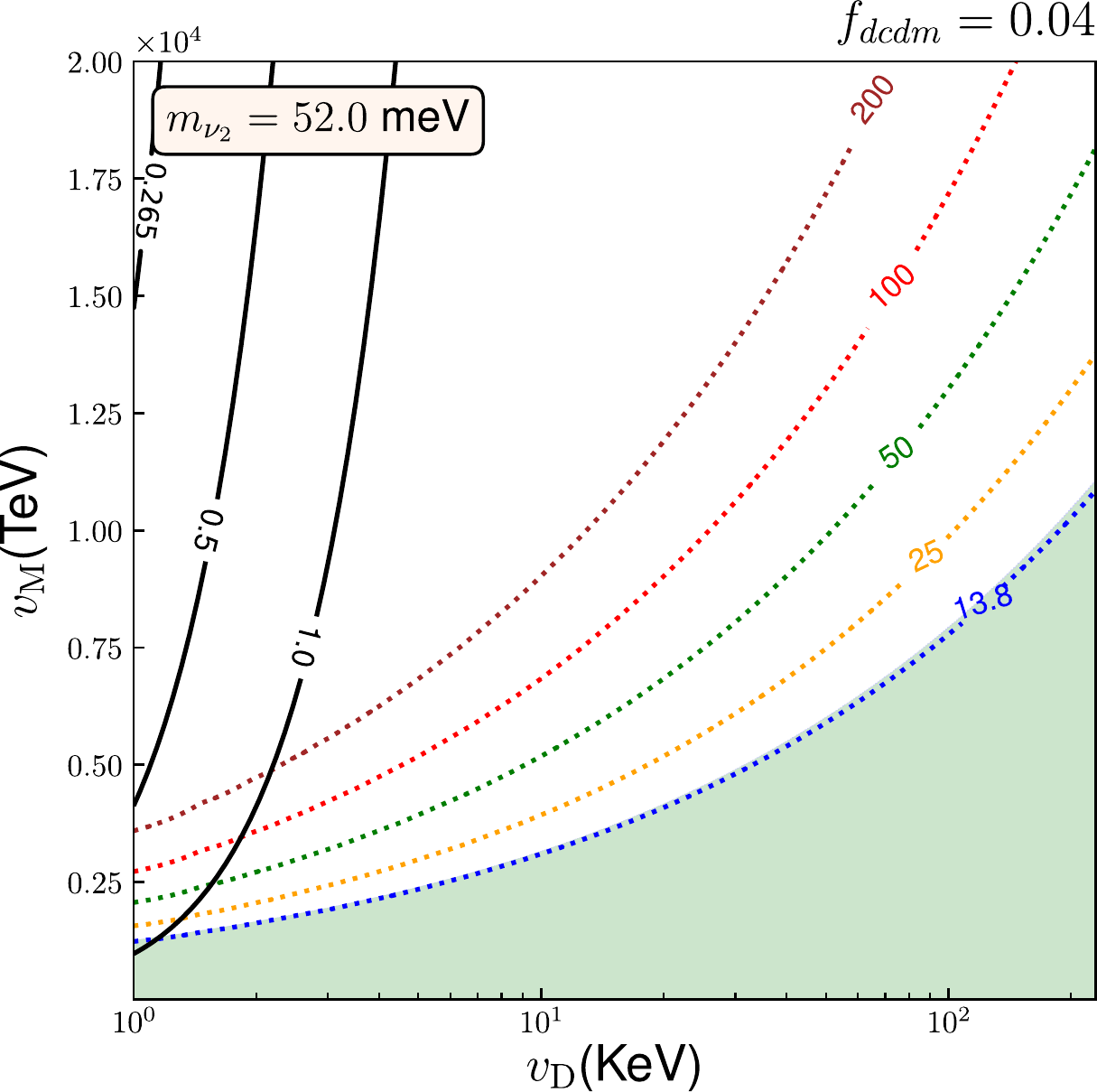}
        \end{subfigure}
        \caption[]{Like Fig.\ref{OmegalifetimeI3NH}, but for neutrino mass inverted hierarchy.}
        \label{OmegalifetimeI3IH}
    \end{figure}

A fraction $f_{\rm dcdm}$ of the initial dark matter could be allowed to decay into dark radiation. Its decay rate $\Gamma_{\rm dcdm}$ has been constrained by a global analysis of both CMB  and low redshift datasets \cite{Audren:2014bca,Poulin:2016nat}. Using these limits, it is possible to put constraints on the new energy scales of our model. Assuming an initial dark matter relative density $\Omega^{\rm ini}_{\rm dm}$ with an initial stable component $\Omega_{\rm sdm}=(1-f_{\rm dcdm})\Omega^{\rm ini}_{\rm sdm}$, we have
 \begin{eqnarray}
\Omega_{\rm dm} &=& \Omega_{\rm sdm} + \Omega_{\rm dcdm} \nonumber \\
            &=& \left[ (1-f_{\rm dcdm}) + f_{\rm dcdm} \, e^{-\Gamma_{\rm dcdm} t}  \right] \Omega^{\rm ini}_{\rm dm}.
\end{eqnarray}
Ref.\cite{Poulin:2016nat} showed that if $\Gamma_{\rm dcdm}< H_0 \sim 0.07$ Gyr$^{-1}$, then $f_{\rm dcdm}\, \Gamma_{\rm dcdm}< \mathcal{O}(10^{-3}){\text{ Gyr}}^{-1}$, where the exact value of the upper bound depends on the dataset. On the contrary, if $\Gamma_{\rm dcdm}> H_0$, then $f_{\rm dcdm} \lesssim 0.04$. We used the most conservative of these bounds to map the parameter space of our model $\vM$ vs. $\vD$, looking for the allowed regions for fixed values of $f_{\rm dcdm}$. Results are shown in Figs. \ref{OmegalifetimeI3NH} and \ref{OmegalifetimeI3IH} where white regions indicate the values of $\Gamma_{I_3}$ compatible with these bounds, blue regions are excluded at 95$\%$ C.L., and green regions are excluded because $\Gamma_{\rm dcdm}> H_0$ and $0.04 \le f_{\rm dcdm}$. Also, black curves indicate different results for the present value of the DM relative density
\begin{equation}
    \Omega_{dm} = \frac{m_{I_3}}{1.25 \text{ KeV}} \, h^{-2} \left( \frac{1-f_{\rm dcdm}}{f_{\rm dcdm}} + e^{-\Gamma_{I_3}t} \right),
\end{equation}
and colored curves represent the different values the lifetime of $I_3$ can take (in Gyr). For example, below the blue dotted curve (13.8 Gyr), $\tau_{I_3}< t_0$ (the age of the universe). The panels on the left and right sides differ in the values of the mass of the lightest (heaviest) neutrino. For instance, in the normal hierarchy, the lightest neutrino is $\nu_1$ and the present limits on its mass are $0 \le m_{\nu_1} \lesssim 30$ meV, whereas in the inverted hierarchy the heaviest neutrino is $\nu_2$ and $49.9$ meV $\lesssim m_{\nu_2} \lesssim 52$ meV. These masses can affect the sum of squared-masses according to the following expression (see Appendix \ref{appendix:sum2}),
\begin{equation}
    \sum_{i=1}^3 m_{\nu_i}^2 = \bigg\{
    \begin{array}{cc}
        3\, m_{\nu_1}^2 + \Delta m_{\rm Sun}^2 + \Delta m_{\rm atm}^2 & (\text{NH}),  \\
        3\, m_{\nu_2}^2 - \Delta m_{\rm Sun}^2 + \Delta m_{\rm atm}^2 & (\text{IH}). 
    \end{array}
    \label{sumMnu2}
\end{equation}
 We calculated the values of $\sum m_{\nu_i}^2$ using Eq.(\ref{sumMnu2}), the central values of $\Delta m_{\rm Sun}^2$ and $\Delta m_{\rm atm}^2$ from \cite{esteban2020fate}, and $\sum m_\nu <$120 meV \cite{pdg}.

\section{Conclusions}
A dark matter model should predict its right abundance today, but this is not enough for a successful model. The model analysed here has some regions on the parameter space that produce the right amount of dark matter, but also present some discrepancies with the standard cosmological model through a very high production of radiation, which would distort the CMB anisotropies, for instance.

The total dark matter of the universe could be multi-component. The lifetime of its decaying component is strongly constrained when the produced dark radiation is formed by neutrinos, gravitational waves or other new relativistic degrees of freedom. In the model presented here, the unstable dark matter candidate $I_3$ decays mainly into neutrinos, allowing us to use the sum of their squared-masses and the fraction $f_{\rm dcdm}$ before decays begin ($\Omega^{\rm ini}_{\rm dcdm} = f_{\rm dcdm} \, \Omega_{\rm dm}^{\rm ini}$) as free parameters to constraint the new energy scales $\vD$ and $\vM$. The latter is the VEV responsible for $B-L$ spontaneous symmetry breaking and the former is responsible for dynamically generating Dirac neutrino masses.

Our results show that, even if taking the most conservative limits on $\Gamma_{\rm dcdm}$, a large area of the parameter space is ruled-out when the decaying DM candidate $I_3$ represents a significant amount of the initial CDM and its total decay rate is lower than the present Hubble parameter, i.e $\Gamma (I_3 \rightarrow \nu \nu)< H_0$, or when the decay rate of $I_3$ is larger than the present value of the Hubble rate and $f_{\rm dcdm} > 0.04$. The strongest bounds come from the present value of the total (stable plus decaying) DM relative density. These are dominant when the initial fraction of decaying dark matter is of a few percent.

\section*{Acknowledgements}

We thank Flavia Sobreira and C\'assio Pigozzo for helpful discussions. GG ackowledges support from CNPq grant No.141699/2016-7 (Brasil), NSERC (Natural Sciences and Engineering Research Council, Canada), McGill Space Institute, and McGill Graduate $\&$ Postdoctoral Studies. PH is partially supported by CNPq (grant no. 310952/2018-2) and FAPESP (grant no. 2014/19164-6). SC is partially supported by CNPq with grant $\#$ 311584/2020-9.

\bibliographystyle{unsrt} 
\bibliography{refs.bib}

\appendix
\section{An approximate expression for the mass of $I_3$}
\label{appendix:mI3}
In order to calculate the mass of the dark matter candidate in this model, it is useful to make the following definitions
\begin{equation}
\zeta \equiv \frac{\vD}{\vM}, \hspace{0.1in} \eta \equiv \frac{\vD}{v_H}, \hspace{0.1in} \chi \equiv \frac{v_H}{\vM},
\label{newparams}
\end{equation}
where $\vD \ll v_H \approx v_{SM} \approx 246$ GeV $\ll \vM$. From Eq.(\ref{VBL}) the mass matrix for CP-odd scalars $M_I$ reads
\begin{equation}
M_I^2 = \frac{v_{SM}^2}{2 \chi^3} \ (1-\eta^2)^2 \ m_I^2,
\end{equation}
where $m_I^2/\chi$ is given by
\begin{equation}
\begin{pmatrix}
(1+\sqrt{2}) \frac{\zeta^2}{\chi}  & \sqrt{2} \ \zeta  & \zeta  & 0 & 0 & 0 & (-2+\sqrt{2}) \ \zeta^2  \\
\sqrt{2} \ \zeta   & \sqrt{2}\chi -\beta_{13} & \beta_{13} & \beta_{13} \zeta & 0 & -\beta_{13} \zeta & \sqrt{2} \zeta \chi  \\
\zeta & \beta_{13}  &  \chi-\beta_{13} & -\beta_{13} \zeta & 0 & \beta_{13} \zeta & -\sqrt{2} \zeta \chi \\
0 & \beta_{13} \zeta & -\beta_{13} \zeta & -\beta_{13} \zeta^2 & 0 & \beta_{13} \zeta^2 & 0 \\
0 & 0 & 0 & 0 &\frac{ 2 \kappa\chi^2}{(1-\eta^2)^2} & 0 & 0\\
0 & -\beta_{13} \zeta & \beta_{13} \zeta & \beta_{13} \zeta^2 & 0 & -9 \beta_{3X} -\beta_{13}\zeta^2 & 3 \beta_{3X} \\
(-2+\sqrt{2}) \ \zeta^2 & \sqrt{2} \zeta \chi & -2 \zeta \chi & 0 & 0 & 3 \beta_{3X} & (4+\sqrt{2}) \zeta^2 \chi-\beta_{3X}
\end{pmatrix}
\label{mI2matrix}
\end{equation}
Since the role of $\eta$ is small, we can safely take $\eta \rightarrow 0$ independently of $\zeta$ and $\chi$. From the $(5,5)$ element of this matrix, we can see one eigenvalue is equal to $2 \kappa \chi^2$ and det($M_I^2$)=0, so at least another one is zero. Solving for their eigenvalues $\lambda$, we get the following eigenvectors in the $\{$ Im(H), Im($\Phi_1$), Im($\Phi_2$), Im($\phi_1$), Im($\phi_2$), Im($\phi_3$), Im($\phi_X$) $\}$ basis
\begin{eqnarray}
\{6 \chi, -9\zeta, 0, -8,0,1,3\} \text{ and } \{-\chi, \zeta, \zeta, 0, 0, 0, 0\} \hspace{0.1in} &\text{for}& \lambda = 0,\\ \nonumber
 \{0,0,0,0,1,0,0\} &\text{for}& \lambda = 2 \kappa \chi^2,
\end{eqnarray}
up to normalization factors. In Ref.\cite{sanchez2014complex} all masses were found (in the limit $\zeta \rightarrow 0$) in the CP-odd scalar sector except for $m_{I_3}$, which was given a numerical estimate of $ \sim \mathcal{O}(\zeta^{1/2})$ in GeV. Here we find an approximate value for this mass, which is the mass of the unstable DM candidate $I_3$. The equation for the eigenvalues of $m_{I}^2$ is
\begin{equation}
\lambda^2 (\lambda - 2 \kappa \chi^3) p(\lambda)=0,
\end{equation}
where $p(\lambda) = \sum_{i=0}^4 c_i \, \lambda^i$ with $c_4$=1. Expanding and comparing coefficients with det($\lambda \, \text{I}_{7\times 7}-m_I^2$), we get (to lowest order in the new parameters defined in Eq.(\ref{newparams})),
\begin{eqnarray}
c_0 & \approx & 74 \sqrt{2} \beta_{13} \beta_{3X} \zeta^2 \chi^6, \nonumber \\
c_1 & \approx & 10\sqrt{2} \chi^5 \beta_{3X} + 11 \sqrt{2} \zeta^2 \chi^5 \beta_{13}-10(1+\sqrt{2})\chi^4 \beta_{13} \beta_{3X},  \nonumber \\
c_2 & \approx & \sqrt{2} \chi^4 -10(1+\sqrt{2})\chi^3 \beta_{3X}-(1+\sqrt{2})\chi^3 \beta_{13} + 20 \chi^2 \beta_{13} \beta_{3X},   \nonumber \\ 
c_3 & \approx & -(1+\sqrt{2})\chi^2+10\chi \beta_{3X} + 2 \chi \beta_{13}.\nonumber 
\end{eqnarray}
As mentioned above, since $m_{I_3}$ is expected to be small, we can expand $p(\lambda)$ around zero $p(\lambda) \approx p(0) + p'(0) \, \lambda$ and solve $p(\lambda)=0$ for $\lambda$, i.e.
\begin{equation}
\lambda \approx -\frac{p(0)}{p'(0)} = - \frac{c_0}{c_1} = \frac{74 \sqrt{2} \beta_{13}\beta_{3X} \zeta^2 \chi^2}{10 \beta_{3X}((1+\sqrt{2})\beta_{13}-\sqrt{2} \chi)-11\sqrt{2} \beta_{13}\zeta^2 \chi},
\end{equation} 
which gives us 
\begin{equation}
m_{I_3}^2 \simeq \frac{v_{SM}^2}{2\chi^3} \lambda = \frac{37\, v_{SM} \, \vM^2 \vD^2 \, \beta_{13}\, \beta_{3X}}{5 \sqrt{2}(1+\sqrt{2}) \vM^3 \beta_{13} \beta_{3X} - v_{SM} \left( 11 \vD^2 \beta_{13} + 10 \vM^2 \beta_{3X} \right)  }.
\end{equation}

\section{Rayleigh-Schrodinger perturbation theory}
\label{appendix:Rayleigh}
Here we find approximate expressions for $I_3$ and its mass using Rayleigh-Schrodinger perturbation theory \cite{lancaster1985theory,mccartin2009rayleigh,Alvarez-Salazar:2019cxw}. First of all, let's write $\tilde{m}_I^2\equiv m_I^2/\chi$ as
\begin{equation}
\tilde{m}_I^2 = \tilde{m}^2_0 + \zeta \tilde{m}^2_1 + \zeta^2 \tilde{m}^2_2,
\end{equation}
where
\begin{equation}
\tilde{m}^2_0= 
\begin{pmatrix}
0 & 0 & 0 & 0 & 0 & 0 & 0  \\
0 & \sqrt{2}\chi - \beta_{13} & \beta_{13} & 0 & 0 & 0 & 0  \\
0 & \beta_{13}  & \chi-\beta_{13} & 0 & 0 & 0 & 0  \\
0 & 0 & 0 & 0 & 0 & 0 & 0  \\
0 & 0 & 0 & 2 \kappa \chi^2 & 0 & 0 & 0  \\
0 & 0 & 0 & 0 & 0 & -9\beta_{3X}  & 3\beta_{3X}   \\
0 & 0 & 0 & 0 & 0 & 3 \beta_{3X}  & -\beta_{3X}   
\end{pmatrix}
\end{equation}

\begin{equation}
\tilde{m}^2_1= 
\begin{pmatrix}
0 & \sqrt{2} & 1 & 0 & 0 & 0 & 0  \\
\sqrt{2} & 0 & 0 & \beta_{13} & 0 & -\beta_{13} & \sqrt{2} \,\chi  \\
1 & 0 & 0 & -\beta_{13} & 0 & \beta_{13} & -2 \, \chi  \\
0 & \beta_{13} & -\beta_{13} & 0 & 0 & 0 & 0  \\
0 & 0 & 0 & 0 & 0 & 0 & 0  \\
0 & -\beta_{13} & \beta_{13} & 0 & 0 & 0 & 0 \\
0 & \sqrt{2} \,\chi & -2 \,\chi & 0 & 0 & 0 & 0   
\end{pmatrix}
\end{equation}

\begin{equation}
\tilde{m}^2_2= 
\begin{pmatrix}
\frac{1+\sqrt{2}}{\chi} & 0 & 0 & 0 & 0 & 0 & -2+\sqrt{2}  \\
0 & 0 & 0 & 0 & 0 & 0 & 0  \\
0 & 0 & 0 & 0 & 0 & 0 & 0  \\
0 & 0 & 0 & -\beta_{13} & 0 & \beta_{13} & 0 \\
0 & 0 & 0 & 0 & 0 & 0 & 0  \\
0 & 0 & 0 & \beta_{13} & 0 & -\beta_{13} & 0 \\
-2+\sqrt{2} & 0 & 0 & 0 & 0 & 0 & (4+\sqrt{2})\chi   
\end{pmatrix}
\end{equation}
The first \textit{unperturbed} eigenvalues and eigenvectors come from $\tilde{m}^2_0$ and they are (before normalization)
\begin{eqnarray}
\lambda^{(0)}_1 &=& 0, \, \lambda^{(0)}_2 = 0, \, \lambda^{(0)}_3 = 0, \, \lambda^{(0)}_4 = -10 \, \beta_{3X}, \,  \lambda^{(0)}_5 = 2 \kappa \chi^2, \nonumber \\ 
 \lambda^{(0)}_{6,7} &=& \left[ -2 \beta_{13} + (1+\sqrt{2}) \chi \mp \sqrt{4\beta_{13}^2 + (3-2\sqrt{2})\chi^2} \right] /2,
\end{eqnarray}

\begin{eqnarray}
I^{(0)}_1 &=& (1,0,0,0,0,0,0), \, I^{(0)}_2 = (0,0,0,1,0,0,0), \, I^{(0)}_3 = (0,0,0,0,0,1,3), \nonumber \\ 
I^{(0)}_4 &=& (0,0,0,0,0,-3,1), \, I^{(0)}_5 = (0,0,0,0,1,0,0), \, I^{(0)}_6 = (0,r^-,1,0,0,0,0), \nonumber \\ 
I^{(0)}_7 &=& (0,r^+,1,0,0,0,0) ,    
\end{eqnarray}
where $2\beta_{13} \, r^\pm = (\sqrt{2}-1)\chi \pm \sqrt{4 \beta_{13}^2 +(3-2\sqrt{2})\chi^2 }$, which is in agreement with \cite{sanchez2014complex}. The first order corrections are obtained using
\begin{equation}
\lambda^{(1)}_i = \left< I^{(0)}_i | \tilde{m}_1^2| I^{(0)}_i \right> \hspace{0.1in} \text{and} \hspace{0.1in} I^{(1)}_i = -\left( \tilde{m}_0^2 - \lambda^{(0)}_i \text{I}_{7\times 7} \right)^{\text{PS}} \left( \tilde{m}_1^2 - \lambda^{(1)}_i \text{I}_{7\times 7} \right) I^{(0)}_i,
\end{equation} 
where I$_{7\times 7}$ is the identity matrix and PS stands for pseudoinverse. For $i=3$ we have $\lambda^{(0)}_3 = 0$, so
\begin{equation}
I^{(1)}_3 = -\tilde{m}_0^{\text{PS}} \tilde{m}_1^2 I^{(0)}_3 = (0,3t-8\beta_{13},-6t+8\sqrt{2}\beta_{13},0,0,0,0),
\end{equation}
where $\sqrt{2} \, t = (2+\sqrt{2})\beta_{13}-2\chi$. Next,
\[
\lambda_3^{(2)} = \left< I^{(0)}_3 | \tilde{m}_0^2| I^{(0)}_3 \right> + \left< I^{(0)}_3 | \tilde{m}_1^2| I^{(1)}_3 \right>  \hspace{4in}
\]
\begin{equation}
\lambda_3^{(2)} = \frac{1}{10}\left( 9(4+\sqrt{2})\chi - \beta_{13} \right) + \frac{1}{3\sqrt{10}N_3} \bigg[ (6t-8\sqrt{2}\beta_{13})(6\chi-\beta_{13})+(3t-8\beta_{13})(3\sqrt{2}\chi-\beta_{13}) \bigg],
\end{equation}
with the normalization factor $N_3=[ 90\chi^2 + 6(2-7\sqrt{2})\beta_{13}\chi + (87-54\sqrt{2})\beta_{13}^2 ]^{1/2}$.

Now, we can compute the second order contribution 
\begin{equation}
I^{(2)}_3 = -(\tilde{m}_0^2)^{\text{PS}} \left[ (\tilde{m}_2^2 - \lambda_3^{(2)})I_3^{(0)}+ \tilde{m}_1^2 I_3^{(1)} \right] = (0,0,0,0,0,-3,1) /\sqrt{10}.
\end{equation}
Finally, $I_3$ and its mass can be approximated to
\begin{equation}
I_3 \simeq I^{(0)}_3 + \zeta \, I^{(1)}_3 + \zeta^2 \, I^{(2)}_3 \hspace{0.2in} \text{and} \hspace{0.2in} m_{I_3}^2 \simeq \frac{v_{SM}^2}{2 \chi^2} \zeta^2 \lambda^{(2)}_3 \approx \frac{1}{2}\vD^2 \lambda_3^{(2)}.
\end{equation}

\section{Sum of neutrino squared masses}
\label{appendix:sum2}
\begin{figure}[t]
\centering
\begin{subfigure}{.5\textwidth}
  \centering
  \includegraphics[width=\textwidth]{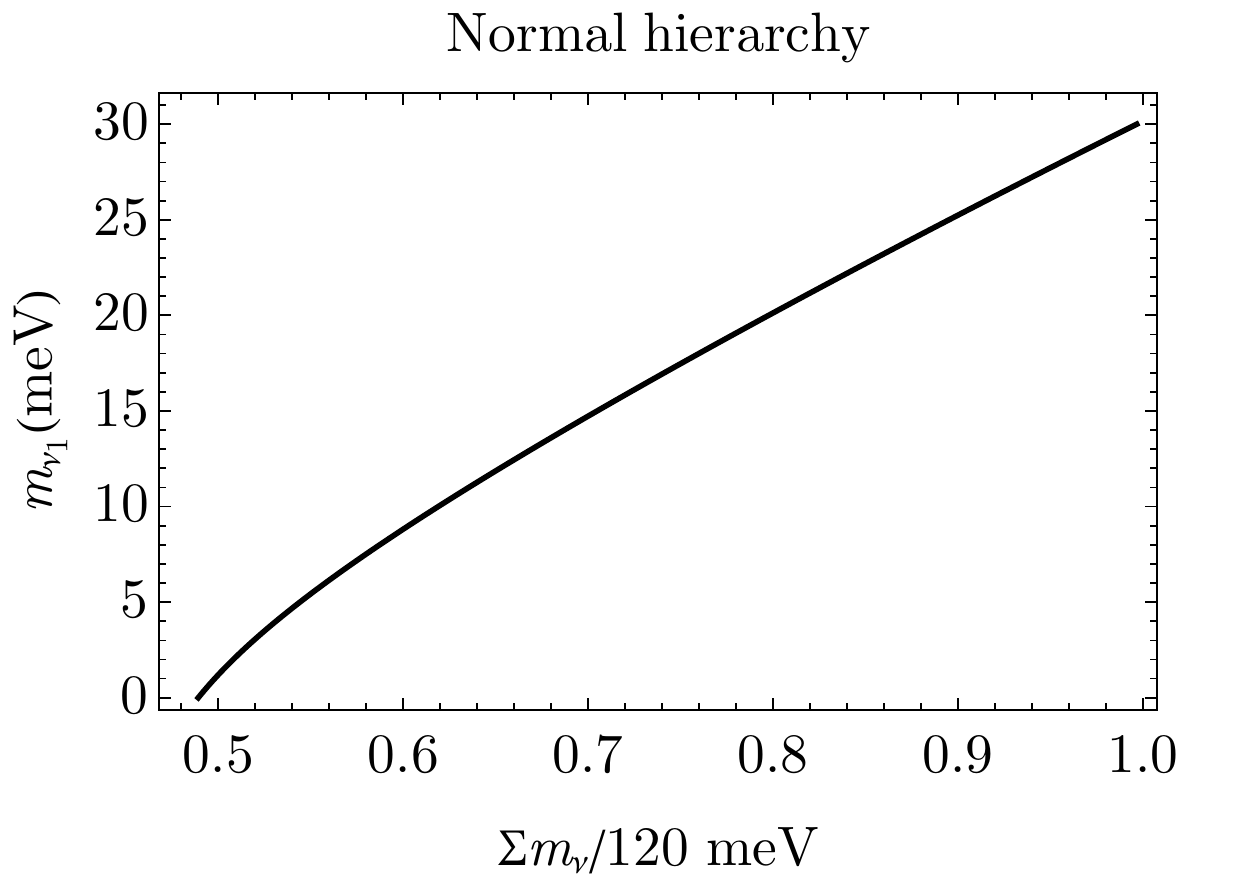}
\end{subfigure}%
\begin{subfigure}{.5\textwidth}
  \centering
  \includegraphics[width=\textwidth]{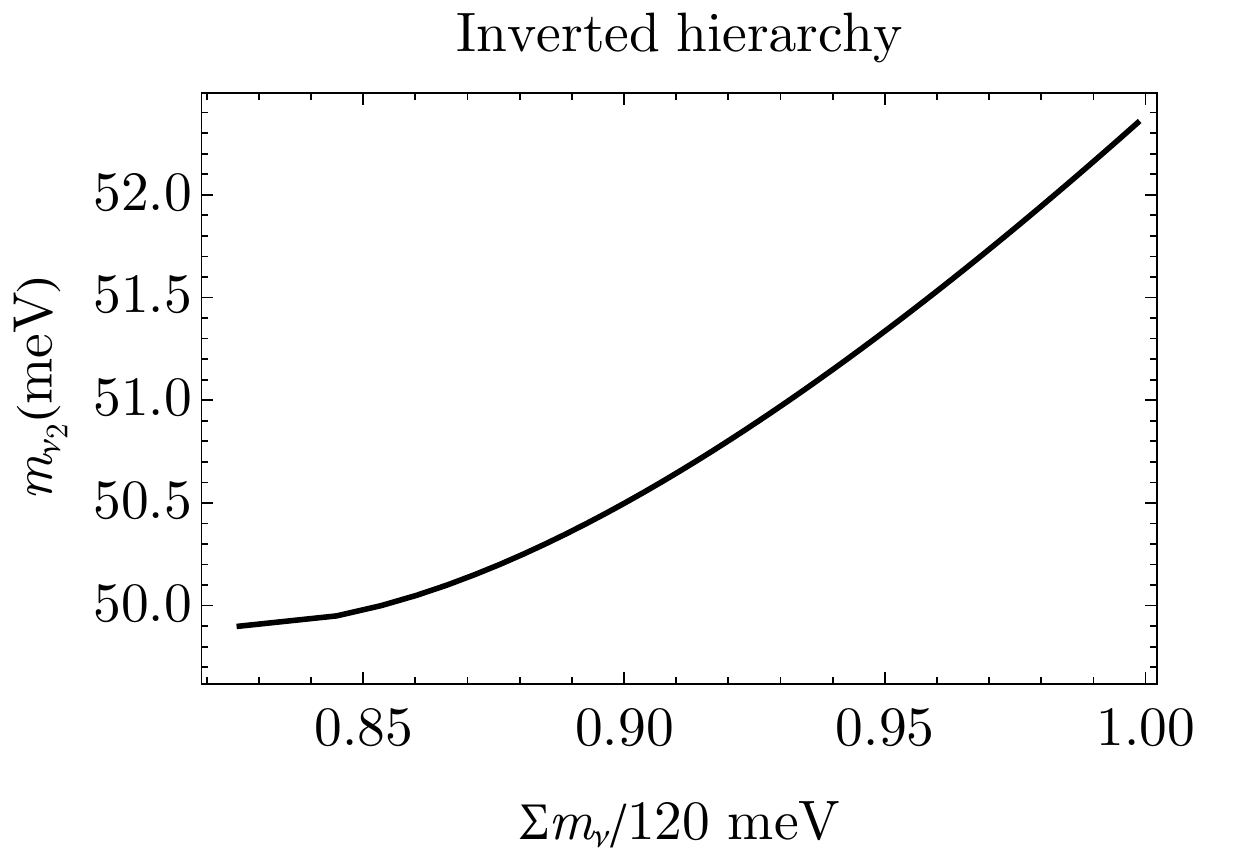}
\end{subfigure}
\caption{Upper and lower bounds on $m_{\nu_1}$ and $m_{\nu_2}$ for the normal and inverted hierarchy, respectively.}
\label{fig:sumNu12}
\end{figure}

\begin{figure}[t]
\centering
\begin{subfigure}{.5\textwidth}
  \centering
  \includegraphics[width=\textwidth]{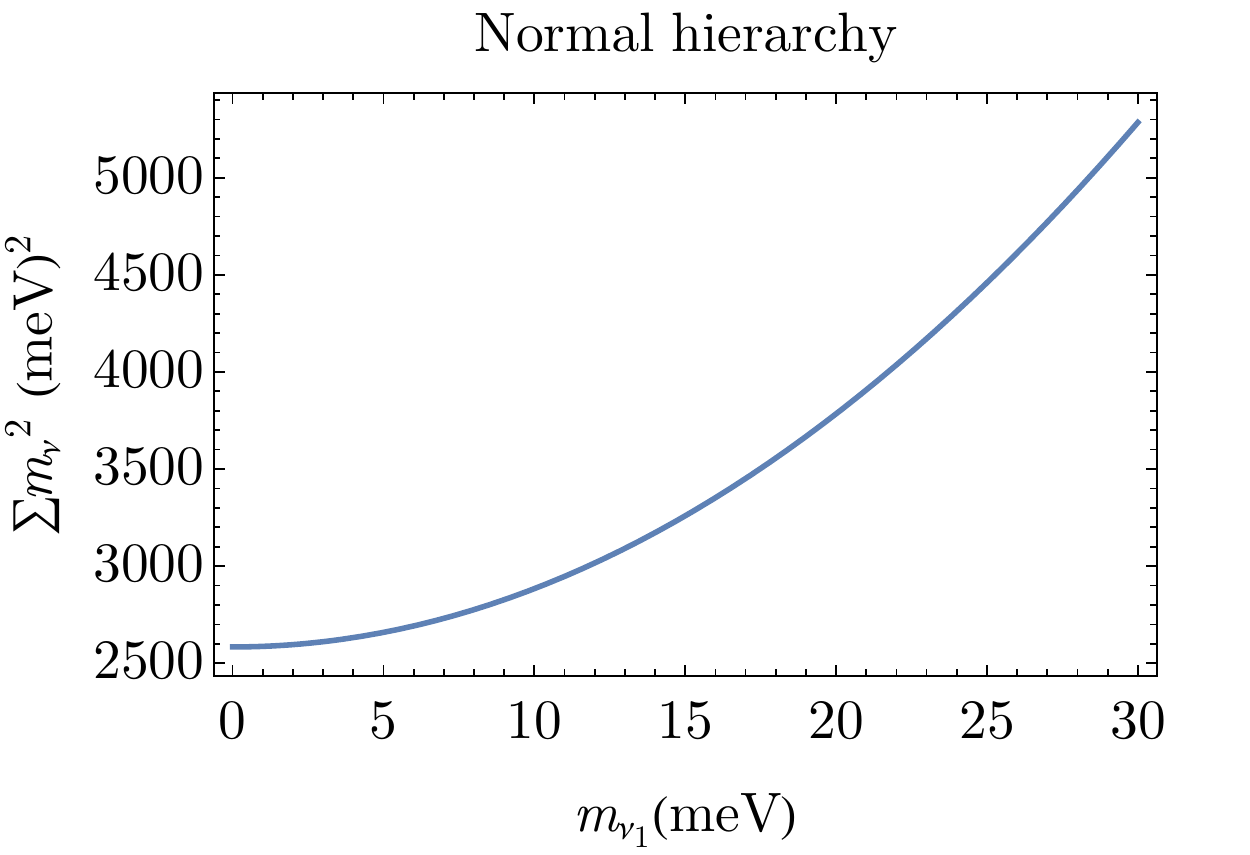}
\end{subfigure}%
\begin{subfigure}{.5\textwidth}
  \centering
  \includegraphics[width=\textwidth]{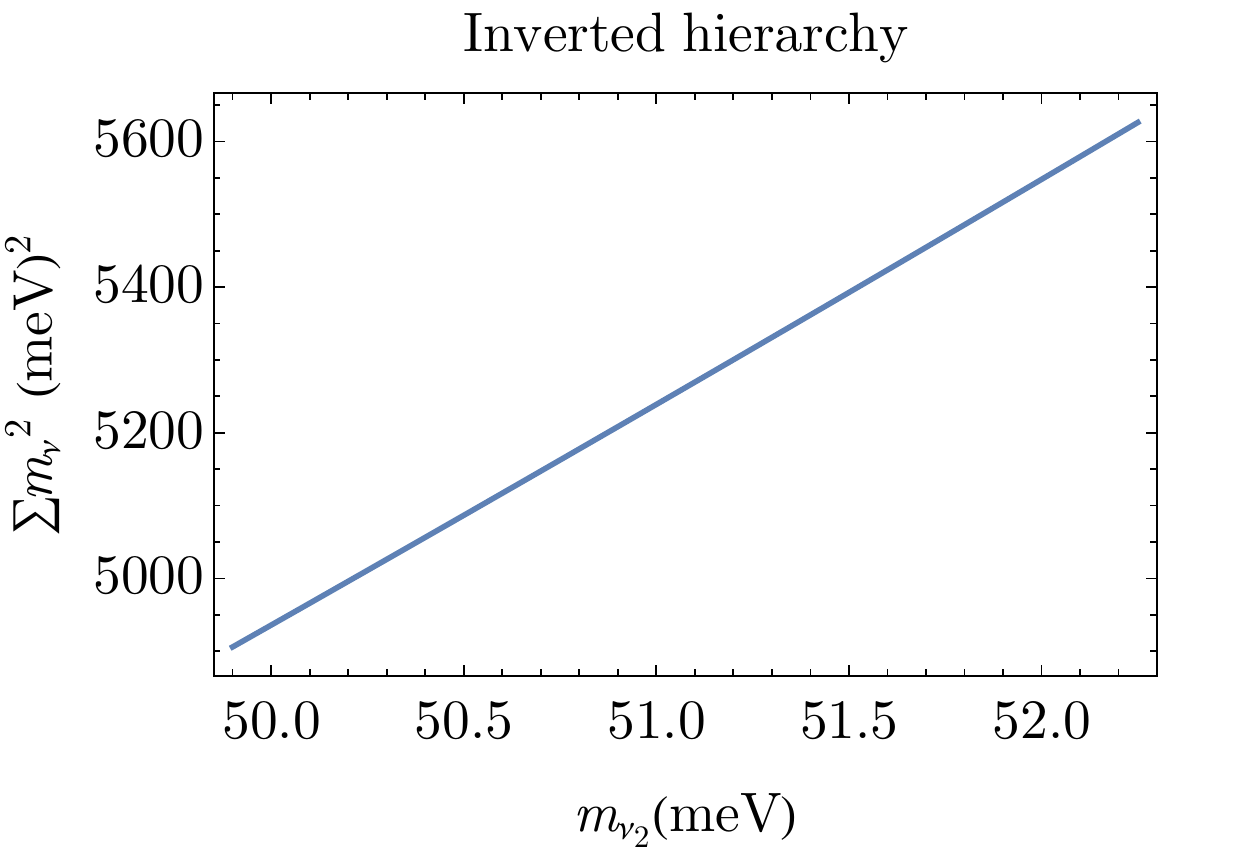}
\end{subfigure}
\caption{Constraints on the sum of neutrino squared masses.}
\label{fig:sumNumasses2}
\end{figure}

When neutrino masses obey a `normal hierarchy' (NH), then $\Delta m_{\rm atm}^2 =  \Delta m_{31}^2$. When the hierarchy is inverted (IH), $\Delta m_{\rm atm}^2 =  \Delta m_{32}^2$. On the other hand, $\Delta m_{\rm Sun}^2 = \Delta m_{21}^2$ irrespective of the hierarchy. Taking this into consideration, we can write the sum of squared masses depending on one neutrino mass only,
\begin{equation}
    \sum_{i=1}^3 m_{\nu_i}^2 = \bigg\{
    \begin{array}{cc}
        3\, m_{\nu_1}^2 + \Delta m_{\rm Sun}^2 + \Delta m_{\rm atm}^2 & (\text{NH})  \\
        3\, m_{\nu_2}^2 - \Delta m_{\rm Sun}^2 + \Delta m_{\rm atm}^2 & (\text{IH}) 
    \end{array}
    \label{sum2}
\end{equation}
The sum of neutrino masses is bounded from above $\sum m_\nu <$ 120 meV \cite{pdg}. Using this limit, $\Delta m_{\rm Sun}^2$, and $\Delta m_{\rm atm}^2$, we get (see Fig.(\ref{fig:sumNu12}))
\begin{eqnarray}
0 \leq &m_{\nu_1} & \lesssim 30 \ {\rm meV} \hspace{0.1in} \text{(Normal hierarchy)}, \nonumber \\
49.9\ {\rm meV} \lesssim & m_{\nu_2} & \lesssim 52.3\  {\rm meV} \hspace{0.1in} \text{(Inverted hierarchy)}.
\end{eqnarray}
Therefore, upper and lower bounds for $\sum m_{\nu_i}^2$ are approximately (see Fig.(\ref{fig:sumNumasses2})) 
\begin{eqnarray}
2500\ {\rm meV^2} \lesssim & \sum m_{\nu_i}^2 & \lesssim 5200\  {\rm meV^2} \hspace{0.1in} \text{(Normal hierarchy)}, \nonumber \\
4900 \ {\rm meV^2} \lesssim & \sum m_{\nu_i}^2 & \lesssim 5600 \ {\rm meV^2} \hspace{0.1in} \text{(Inverted hierarchy)}.
\end{eqnarray}

\end{document}